\documentclass[ reprint, floatfix, superscriptaddress, amsmath, amssymb, longbibliography ]{revtex4-1}
 \pdfoutput=1
\usepackage[utf8]{inputenc}
\usepackage[T1]{fontenc}
\usepackage{graphicx}   
\usepackage{dcolumn}    
\usepackage{bm}         
\usepackage{verbatim}   
\usepackage{booktabs}
\usepackage{footnote}
\usepackage{subfigure}
\usepackage{xcolor}
\usepackage{float}
\usepackage{amsmath}
\usepackage[most]{tcolorbox}
\usepackage{soul}
\usepackage{mathtools}
\usepackage{siunitx}
\usepackage{braket}
\usepackage{circuitikz}
\usepackage{tikz}
\usetikzlibrary{arrows.meta, positioning, calc, decorations.pathmorphing, shapes.multipart}
\usepackage{blkarray}
\usepackage[colorlinks=true,linkcolor=blue,allcolors=blue]{hyperref}
\newcommand{\figref}[1]{\mbox{Fig.~\ref{#1}}}

\newcommand{\figpanel}[2]{Fig.~\hyperref[#1]{\ref*{#1}(#2)}}
\newcommand{\figpanels}[3]{Fig.~\hyperref[#1]{\ref*{#1}(#2)--(#3)}}
\newcommand{\figpanelNoPrefix}[2]{\hyperref[#1]{\ref*{#1}(#2)}}

\renewcommand{\eqref}[1]{\mbox{Eq.~(\ref{#1})}}
\usepackage{mleftright} 
\usepackage{url}

\begin{document}

\author{Mustafa Bakr}
 \email{mustafa.bakr@physics.ox.ac.uk} 
 \affiliation{Department of Physics, Clarendon Laboratory, University of Oxford, OX1 3PU, UK}
 \author{Mohammed Alghadeer}
 \affiliation{Department of Physics, Clarendon Laboratory, University of Oxford, OX1 3PU, UK}
\author{Simon Pettersson Fors}
\affiliation{Department of Microtechnology and Nanoscience, 
Chalmers University of Technology, 412 96 Gothenburg, Sweden} 
\author{Simone D. Fasciati}
 \affiliation{Department of Physics, Clarendon Laboratory, University of Oxford, OX1 3PU, UK}
\author{Shuxiang Cao}
 \affiliation{Department of Physics, Clarendon Laboratory, University of Oxford, OX1 3PU, UK}
\author{Atharv Mahajan}
 \affiliation{Department of Physics, Clarendon Laboratory, University of Oxford, OX1 3PU, UK}
\author{Smain Amari}
 \affiliation{Department of Physics, Clarendon Laboratory, University of Oxford, OX1 3PU, UK}
\author{Anton Frisk Kockum}
\affiliation{Department of Microtechnology and Nanoscience, 
Chalmers University of Technology, 412 96 Gothenburg, Sweden}
\author{Peter Leek}
\affiliation{Department of Physics, Clarendon Laboratory, University of Oxford, OX1 3PU, UK}

\title{Intrinsic Multi-Mode Interference for Passive Suppression of Purcell Decay in Superconducting Circuits}

\begin{abstract}
Decoherence due to radiative decay remains an important consideration in scaling superconducting quantum processors. We introduce a passive, interference-based methodology for suppressing radiative decay using only the intrinsic multi-mode structured environment of superconducting circuits. By taking into account the full electromagnetic mode-mode couplings within the device, we derive analytic conditions that enable destructive interference. These conditions are realized by introducing controlled geometric asymmetries—such as localized perturbations to the transmon capacitor—which increase mode hybridization and activate interference between multiple decay pathways. We validate this methodology using perturbation theory, full-wave electromagnetic simulations, and experimental measurements of a symmetry-broken transmon qubit with improved coherence times.
\end{abstract}


\maketitle


\section{Introduction}
The implementation of fault-tolerant quantum information processing~\cite{ref1, ref2, ref3} demands high-fidelity quantum gates alongside fast and accurate qubit-state measurement. In circuit quantum electrodynamics (cQED)~\cite{ref4, ref5, ref6}, the qubit state is typically measured by detecting the state-dependent frequency shift of a readout resonator via homodyne detection~\cite{ref7, ref8, ref9}. However, this measurement approach introduces an undesired decay channel for the qubit known as the Purcell effect~\cite{ref10}, arising from energy leakage into the readout lines. This radiative decay significantly limits qubit coherence and measurement fidelity. 

The Purcell decay rate for a qubit coupled to a resonator in the dispersive regime depends on three key factors: the qubit-resonator coupling strength, the resonator decay rate, and the frequency detuning between qubit and resonator~\cite{ref11}. Consequently, there exists a fundamental trade-off: rapid measurement requires strong coupling and broad resonator bandwidth, while prolonged qubit coherence necessitates the opposite. This challenge has motivated multiple approaches to navigate this trade-off and reduce Purcell decay while preserving fast and efficient measurement.

Existing solutions include the use of Purcell filters~\cite{ref12, ref13, ref14}, engineering Purcell-protected qubits~\cite{ref15, ref16, ref17}, and implementing tunable couplers~\cite{ref18}. Purcell filters are designed to impede photon propagation at the qubit frequency while allowing transmission at the readout frequency. These include notch filters that create a rejection band around the qubit frequency~\cite{ref19}, and bandpass filters that restrict transmission to a narrowband around the resonator frequency~\cite{ref20, ref21, ref22, ref23}. The bandpass approach offers the additional advantage of protecting qubits with a range of frequencies and enabling frequency-multiplexed readout of multiple qubits~\cite{ref24, ref25, ref26}. While these techniques have proven effective, they share a common limitation: they rely on adding external circuit elements to suppress the Purcell decay. As quantum processors scale up to hundreds and thousands of qubits, the overhead associated with incorporating individual filter structures becomes increasingly problematic in terms of chip area, design complexity, and potential failure points. 

In this article, we propose a fundamentally different approach to suppress Purcell decay, that requires no additional circuit elements. We exploit the intrinsic multi-mode electromagnetic environment naturally present within superconducting circuits to engineer destructive interference among multiple decay pathways. Our central insight is that qubit-resonator systems naturally support a spectrum of high-frequency modes beyond their fundamental frequencies typically considered in simplified models. By introducing controlled asymmetry in the circuit geometry — such as localized mouse-bite perturbations~\cite{ref27, ref28, ref29} or slots in the transmon qubit's capacitor pads — we activate couplings between otherwise decoupled modes, creating multiple interfering emission paths. 

The interference mechanism differs conceptually from traditional filtering approaches. Rather than blocking the qubit's emission after it has coupled to a resonator mode, we engineer the intrinsic electromagnetic environment to create interference between multiple emission pathways. This is analogous to interference effects in classical electrodynamics~\cite{ref30}, atomic physics~\cite{ref31} or giant atoms in waveguide quantum electrodynamics~\cite{ref32}. In both simulation and measurement, we demonstrate that breaking the symmetry of this distributed structure can significantly enhance qubit relaxation times via multi-mode interference. In fabricated devices with engineered capacitor asymmetry, we observe clear signatures of Purcell suppression at interference sweet spots with a factor of two improvement in  qubit relaxation times, consistent with both our analytic theory and numerical models. We anticipate that this method will be broadly applicable to a wide range of superconducting qubit architectures. Future extensions may exploit this framework for combined suppression of radiative loss and ZZ-coupling errors~\cite{ref33, ref34, ref35, ref36, ref37}, potentially enabling fixed or tunable protection mechanisms based on interference for error mitigation.

This article is organized as follows. Section~\ref{multimodecircuit} introduces the multi-mode circuit model. Section~\ref{sec:AnalysisSingleExc} details our semiclassical and quantum analysis of interference-based Purcell suppression in the single-excitation subspace. Section~\ref{sec:AnalysisDrivenPurcell} extends our analysis to the driven case. Section~\ref{sec:Simulation} presents numerical simulations, and Section~\ref{sec:Experiment} experimental results, that confirm our theoretical predictions. We conclude in Section~\ref{sec:Discussion}, discussing implications for scalable superconducting quantum architectures.

\begin{figure*}
\centering
\includegraphics[width=0.75\textwidth]{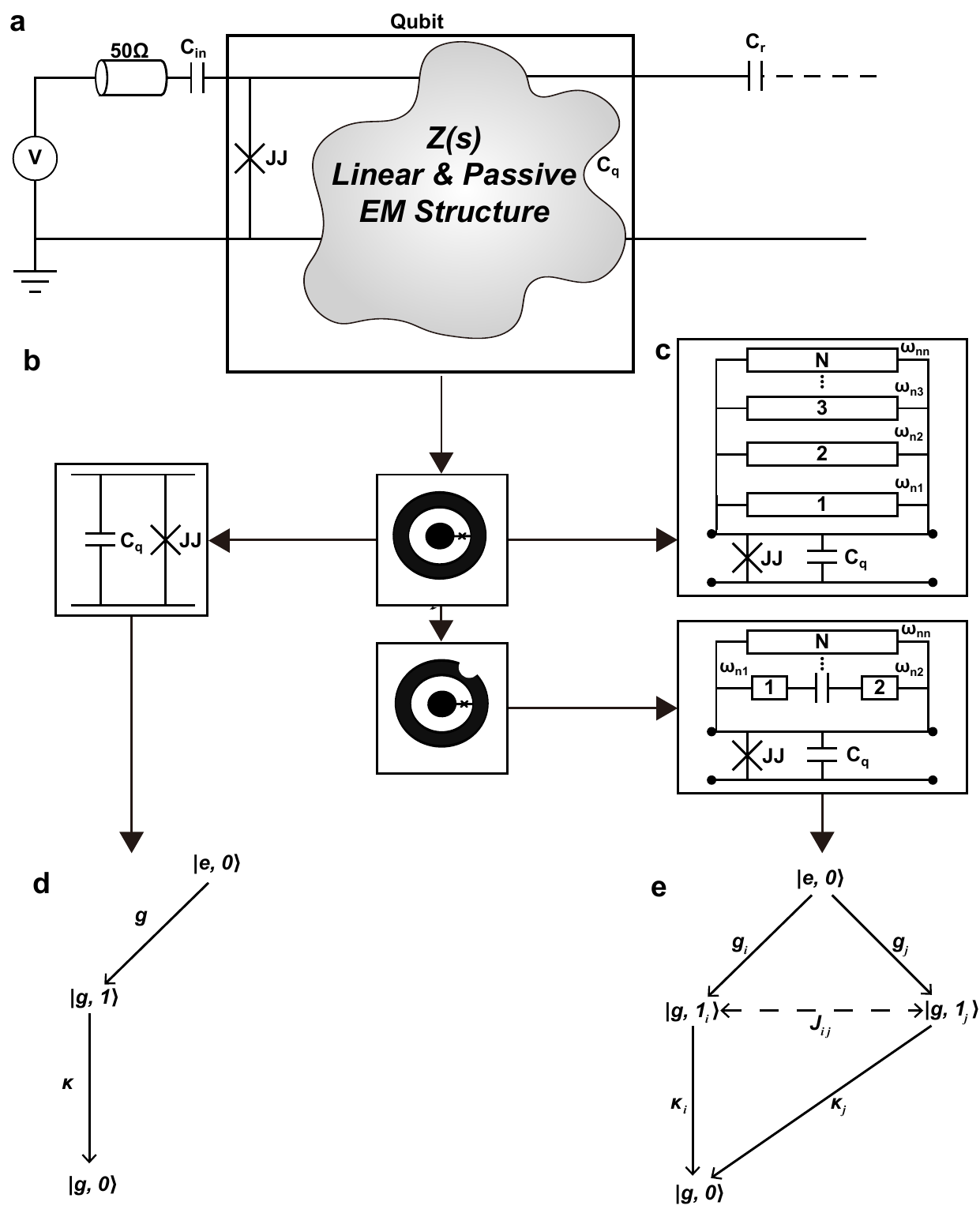}
\caption{(a) Conceptual illustration of the multi-mode nature of a transmon circuit, where the shunt capacitor is replaced by a generalised electromagnetic (EM) structure \( Z(s) \) representing the distributed nature of the qubit. Here, \( C_\text{in} \) denotes the input coupling capacitor, \( C_q \) the qubit shunt capacitance, JJ the Josephson junction, and \( C_r \) the coupling capacitance to the readout resonator. (b) Traditionally, the transmon is modelled as a lumped-element circuit comprising a Josephson junction in parallel with a single shunt capacitor \( C_q \), representing a single quantised mode. (c) Multi-mode circuit representation of the large qubit capacitor, including higher electromagnetic modes and their possible coherent couplings $J_{ij}$. The top diagram shows the idealized symmetric case: the large capacitor supports a ladder of orthogonal, nominally uncoupled modes at frequencies $\omega_{n1}, \omega_{n2}, \ldots, \omega_{nn}$, arising from its rotational symmetry. In the lower diagram, a small geometric perturbation—such as a ``mouse-bite'' cut in the capacitor pad—breaks this symmetry and hybridizes otherwise orthogonal modes. This enables coupling between near-degenerate modes, transforming the topology from parallel (transverse) to coupled (longitudinal) mode configurations with effective inter-mode couplings. (d) Quantum state transitions in the traditional single-mode model. States are denoted 
$\lvert q, n \rangle$, where $q \in \{g, e\}$ indicates the qubit ground or excited state, and $n = 0, 1$ is the photon number in the resonator. A single decay path exists via the qubit–resonator coupling $g$, followed by resonator photon loss at rate $\kappa$. (e) Quantum state transitions in the multi-mode model. Two hybridized decay channels through $\lvert g, 1_i \rangle$ and $\lvert g, 1_j \rangle$ are coupled through coherent mode–mode interaction $J_{ij}$, modifying the effective radiative decay. Solid arrows represent qubit–mode couplings $g_i$, $g_j$ and photon loss rates $\kappa_i$, $\kappa_j$; the dashed arrow indicates coherent mode–mode coupling.
}
\label{fig:multimode_concept}
\end{figure*}

\section{Multi-mode circuit fundamentals}
\label{multimodecircuit}

Superconducting qubits, such as the transmon~\cite{ref38}, are commonly modeled as lumped-element circuits comprising one or more Josephson junctions (JJs), depending on their frequency tunability and design architecture. Figure~\figpanelNoPrefix{fig:multimode_concept}{a, b} illustrate that, while this simplified picture captures the essential anharmonicity and enables tractable single-mode analyses, it fails to reflect the distributed electromagnetic structure inherent to the physical geometry of these devices. In practice, the large transmon shunting capacitor pads and surrounding layout act as extended resonant structures, giving rise to a rich set of higher modes that participate in the circuit dynamics~\cite{ref39}, illustrated in \figpanel{fig:multimode_concept}{c, e}. In this section, we capture the multi-mode nature of superconducting quantum circuits.
\subsection{From single-mode to multi-mode description}
The transmon is conventionally treated as a weakly anharmonic oscillator with a single electromagnetic mode~\cite{ref40}. In cQED architectures, a qubit is typically coupled  dispersively to a single readout resonator mode~\cite{ref5}. This interaction is described by the Jaynes–Cummings Hamiltonian $(\hbar = 1$)
\begin{multline}
H_{JC} = \omega_q \frac{\sigma_z}{2} + \omega_r a^\dagger a 
+ g(\sigma_+ a + \sigma_- a^\dagger) \\
+ \Omega_q\mleft( e^{i\omega_d t} \sigma_+ + e^{-i\omega_d t} \sigma_- \mright) 
+ \epsilon_r\mleft( e^{i\omega_p t} a + e^{-i\omega_p t} a^\dagger \mright),
\end{multline}
where $\omega_q$ and $\omega_r$ are the qubit and resonator frequencies, respectively, $g$ is their coupling strength, and $\Omega_q$ and $\epsilon_r$ represent external drive amplitudes at frequencies $\omega_d$ and $\omega_p$, respectively. Here, \( a \) and \( a^\dagger \) are the annihilation and creation operators for the resonator mode, while \( \sigma^- \), \( \sigma^+ \), and \( \sigma_z \) are the lowering, raising, and Pauli-\( z \) operators for the qubit. This model is approximate, but has proven successful in describing dispersive readout~\cite{ref41}, gate operations, and energy
relaxation via the Purcell effect. 

This single-mode approximation, however, fails to account for the complex electromagnetic environment arising from the extended geometry of the circuit elements. The physical implementation of a transmon involves spatially distributed capacitor pads, and similarly, readout resonators possess complex geometries that support multiple electromagnetic modes. Full-wave electromagnetic simulations of actual layouts using finite-element solvers (e.g., Ansys HFSS) reveal that even simple circuit geometries support a spectrum of self-resonances beyond their fundamental modes. These high-frequency modes are typically neglected in conventional analyses, but they provide additional degrees of freedom that can be harnessed for coherence engineering. Figure~\ref{fig:multimode_concept} illustrates the conceptual difference between the traditional single-mode model and our multi-mode description. For example, the eigenmode frequencies obtained from full-wave simulations of the transmon's capacitive pads described in Ref.~\cite{ref42} reveal higher modes at $f_{2v}^q$ = \SI{45.2}{\giga\hertz} and $f_{2h}^q$ = \SI{45.2}{\giga\hertz}, corresponding to vertical and horizontal charge oscillations, respectively. Similarly, a spiral readout resonator with a fundamental frequency at $f_0 = \SI{9.1}{\giga\hertz}$ exhibits additional resonances at $f_{1r} = \SI{21.9}{\giga\hertz}$ and $f_{2r} = \SI{35.7}{\giga\hertz}$.

To comprehensively model these interactions, we extend the Hamiltonian to include $m$ interacting modes and their mutual couplings
\begin{multline}
H = \frac{\omega_q}{2}\sigma_z + \sum_{i=1}^{m}\omega_i a_i^\dagger a_i 
+ \sum_{i=1}^{m} g_i\mleft(e^{i\phi_i}a_i\sigma_+ + e^{-i\phi_i}a_i^\dagger\sigma_-\mright)
+ \\ \sum_{i \neq j}^{m} J_{ij}e^{i\theta_{ij}}a_i^\dagger a_j 
+ \Omega_q\mleft(e^{i\omega_d t}\sigma_+ + e^{-i\omega_d t}\sigma_-\mright) 
+ \\\sum_{i=1}^{m}\epsilon_{i}\mleft(e^{i\omega_{pi}t}a_i + e^{-i\omega_{pi}t}a_i^\dagger\mright),
\label{eq:multi_mode_H}
\end{multline}
where $J_{ij}$ and $\theta_{ij}$ describe the coupling strength and phase between modes $i$ and $j$, and $\phi_i$ accounts for the phase of each mode at the qubit's location. The coefficients \( g_i \) denote the coupling strength between the qubit and mode \( i \), while \( \epsilon_i \) represents the external drive amplitude on mode \( i \). Note that $\theta_{ij} = -\theta_{ji}$ is antisymmetric due to the fact that the Hamiltonian is Hermitian. These phases arise from the physical geometry of the circuit and the spatial distribution of electromagnetic fields~\cite{ref43}. We note that this Hamiltonian coincides with that of a ``giant atom'' coupled at multiple points to a structured environment formed by coupled resonators~\cite{Soro2023, Ingelsten2024}.

Importantly, in structures with perfect spatial or electromagnetic symmetry—such as mirror symmetry about the mode coupling interfaces—the inter-mode coupling strengths \( J_{ij} \) vanish. However, when this symmetry is deliberately broken, for example by geometric perturbations or asymmetric placement of ports or junctions, \( J_{ij} \) becomes finite, thereby enabling the interference pathways central to our approach.

\subsection{Symmetry breaking and mode hybridization}

The geometric symmetry in circuit design strongly influences the coupling between different electromagnetic modes. In perfectly symmetric transmon geometries, the fundamental qubit mode and higher modes remain decoupled due to orthogonal field distributions. However, introducing asymmetries arising from fabrication defects or intentional perturbations—such as asymmetric gaps, off-center vias, or localized notches (sometimes referred to as ``mouse bites'')—can hybridize otherwise orthogonal modes. This results in finite off-diagonal coupling elements between high-frequency modes and between these modes and the qubit, leading to a more complex environment for the qubit. The significance of this effect can be quantified through energy participation ratios (EPRs)~\cite{ref44}, which measure the degree to which each mode couples to the Josephson junction. 

\begin{figure}
\centering
   \includegraphics[width=0.75\linewidth]{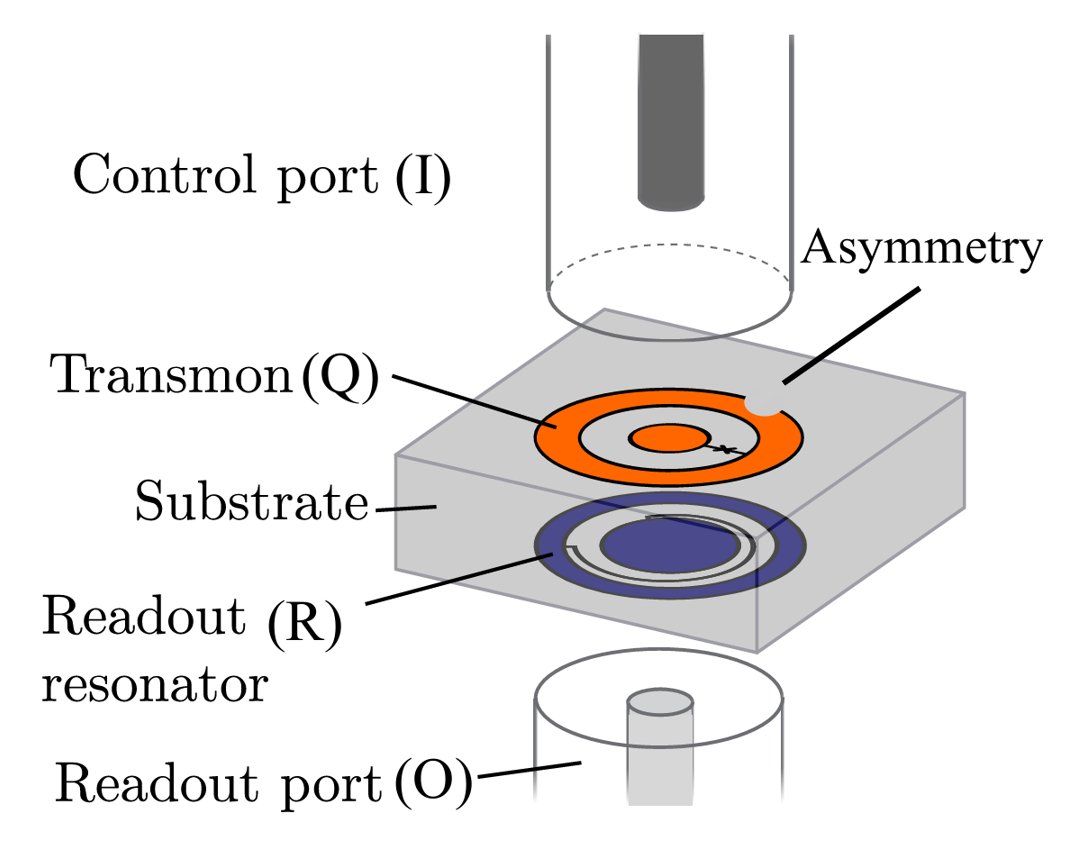}
   \caption{Device schematic and simulation setup. A simplified 3D schematic illustrates the coaxial pad geometry of the device, including an asymmetry-breaking feature (in this case, a mouse bite) in the outer ring of the transmon capacitor. The labeled ports—control (I), transmon (Q), readout resonator (R), and readout output (O)—correspond to voltage boundaries defined in the HFSS Maxwell solver for calculating the electrostatic capacitance matrix. The metallic enclosure, which acts as the global ground node (Gnd), is omitted here for visual clarity. 
}
   \label{fig:circuit1}
\end{figure}

We simulate the example shown in \figref{fig:circuit1} and compare the EPRs in symmetric and symmetry-broken configurations. The junction’s participation in the fundamental qubit mode remains essentially unchanged at 0.97 in both layouts. However, introducing symmetry breaking induces significant increases in participation from other modes. For instance, for the transmon capacitor pads mode at \( f_{2v}^q = \SI{45.1}{\giga\hertz} \), the EPR increases from \( 3.5 \times 10^{-5} \) to \( 1.5 \times 10^{-3} \); and at \( f_{2h}^q = \SI{45.3}{\giga\hertz} \), from \( 5.1 \times 10^{-6} \) to \( 3.7 \times 10^{-4} \). These shifts represent enhancements of about two orders of magnitude, confirming increased mode hybridization, which could enable interference effects central to our suppression mechanism.

We further simulate the electrostatic behavior of the circuit shown in \figref{fig:circuit1} using Ansys Maxwell. While the physical layout of both the transmon qubit and the resonator consists of two floating capacitor pads each~\cite{ref45, ref46}, we model them in the extracted capacitance matrix using simplified lumped elements labeled as “Q” (qubit) and “R” (resonator), respectively. That is, we reduce each floating structure to a single node to capture the dominant mode of interaction while omitting intra-structure effects. This simplification is made for consistency with our equivalent circuit model. To study the effect of geometric asymmetries, we perform two separate electrostatic simulations: one with ideal mirror symmetry, and one with an intentionally broken geometry introduced by a small ``mouse bite'' feature in the transmon outer capacitor pad. Let \( C_{\text{sym}} \) denote the capacitance matrix extracted from the symmetric geometry, and \( C_{\text{asym}} \) from the asymmetric geometry. We then define the relative difference matrix as \( \Delta C = (C_{\text{asym}} - C_{\text{sym}}) / C_\text{sym} \), which quantifies the additional capacitive coupling pathways activated by the symmetry-breaking perturbation. The control and readout ports are labeled “I” (input) and “O” (output), and “Gnd” represents the global ground reference. The resulting difference matrix is
\begin{equation}
\Delta C~[\text{pF}] =
\begin{pmatrix}
\text{} & \text{Gnd} & \text{I} & \text{O} & \text{Q} & \text{R} \\
\text{Gnd} & 0.992 & 1.000 & 1.000 & 0.909 & 1.016 \\
\text{I}    & 1.000 & 1.000 & 35.341 & 0.965 & 39.253 \\
\text{O}   & 1.000 & 35.341 & 1.000 & 0.934 & 1.146 \\
\text{Q}   & 0.916 & 0.964 & 0.934 & 0.918 & 0.928 \\
\text{R}   & 1.019 & 39.008 & 1.167 & 0.927 & 0.992 
\end{pmatrix}
.
\label{eq:deltaC_matrix}
\end{equation}
Values greater than unity indicate an increase in mutual capacitance due to symmetry breaking, while values less than unity imply suppression. These shifts confirm that even small geometric asymmetries can introduce non-negligible cross-capacitance between qubit, resonator, and control ports. As a result, the electromagnetic mode hybridization is modified, potentially activating new multi-path decay or interference channels within the circuit.

The central objective of this article is to explore how this rich multi-mode structure---when properly characterized and controlled---can be used to suppress qubit decay through destructive interference among multiple radiative channels. In particular, we show that intrinsic multi-mode interference, arising from the coherent superposition of amplitudes in the extended Hamiltonian described above, can be engineered to cancel the leading-order radiative decay terms by leveraging weak couplings to high-frequency modes. These couplings can be controlled through geometric design.  In the following section, we develop an analytical model from a semi-classical description of this interference effect.

\section{Analysis of the interferometric Purcell filter in the single-excitation subspace}
\label{sec:AnalysisSingleExc}

To quantitatively analyze the effect of intrinsic multi-mode interference on radiative decay, we begin with a semiclassical treatment of the qubit’s dynamics in the single-excitation regime, where the qubit has at most one excitation and the mode populations remain near vacuum. Our aim is to derive an analytic expression for the effective decay rate (Purcell rate) that includes phase-dependent interference between all decay channels mediated by the internal electromagnetic modes of the system. We consider a qubit coupled to \( m \) modes, isolate the passive decay dynamics, and set all external drive amplitudes in \eqref{eq:multi_mode_H} to zero, i.e., \( \Omega_q = 0 \) and \( \epsilon_{i} = 0 \). 

We work in the lab frame initially, and use the Heisenberg-Langevin equation for \( \sigma^- \),
\begin{equation}
\dot{\sigma}^{-} = -i\omega_q \sigma^{-} - i \sum_{i=1}^{m} g_i e^{i\phi_i} a_i ,
\label{eq:qubit_motion}
\end{equation}
where $\sigma_-$ represents the qubit lowering operator and $g_i$ denotes the coupling strength between the qubit and the $i$th mode. The operators $a_i$ and $a_j$ represent resonator field amplitudes. Similarly, for each mode \( a_i \), incorporating energy loss via the resonator decay rate \( \kappa_i \), we obtain the corresponding Heisenberg–Langevin equation of motion,
\begin{equation}
\dot{a}_i = \mleft(-i\omega_i - \frac{\kappa_i}{2} \mright) a_i - i g_i e^{-i\phi_i} \sigma^{-} - i \sum_{j \neq i}^m J_{ij} e^{i\theta_{ij}} a_j,
\label{eq:resonator_motion}
\end{equation}
where we recall $\theta_{ij} = - \theta_{ji}$. Since we are interested in how the resonators interacts with the qubit, it is convenient to rewrite the equation in a frame rotating at the qubit frequency $\omega_q$, where the detuning is
\begin{equation}
\Delta_i = \omega_q - \omega_i.
\label{eq:detuning}
\end{equation}
Rewriting \eqref{eq:resonator_motion} in the rotating frame, using \( a_i(t) \rightarrow a_i(t)e^{-i\omega_q t} \), we obtain
\begin{equation}
\dot{a}_i = \mleft(i\Delta_i - \frac{\kappa_i}{2} \mright) a_i - i g_i e^{-i\phi_i} \sigma^{-} - i \sum_{j \neq i}^m J_{ij} e^{i\theta_{ij}} a_j.
\label{eq:simplified_resonator_mode}
\end{equation}

Under a Born-Oppenheimer approximation, we assume the resonator modes reach their steady state much faster than the qubit evolves ($\kappa_i \gg$ effective qubit decay rate). This separation of timescales allows us to set $\dot{a}_i = 0$ while still allowing $\sigma^-$ to evolve. In physical terms, this means the resonator field adjusts instantaneously to changes in the qubit state because the resonator's characteristic response time ($1/\kappa_i$) is much shorter than the qubit's evolution timescale. This approximation allows us to express the resonator amplitudes in terms of the instantaneous qubit state. Solving for $a_i$ then yields
\begin{equation}
a_i = \frac{ -ig_i e^{-i\phi_i} \sigma^{-} -i  \sum_{j \neq i}^m J_{ij} e^{i\theta_{ij}} a_j }{ i\Delta_i +  \kappa_i / 2 }.
\label{eq:steady_state_a}
\end{equation}

Substituting \eqref{eq:steady_state_a} into \eqref{eq:qubit_motion} [noting that \eqref{eq:qubit_motion} must first be transformed into the rotating frame of the qubit], and truncating to first order in \( J_{ij} \)—i.e., assuming \( \sum_{j \neq i} |J_{ij}| \ll |\Delta_i| \) for all modes \( i \),
and inserting only the zeroth-order \( a_j \) values into the cross-term (see Appendix~\ref{appendixB})—we obtain
\begin{align}
\dot{\sigma}^{-} = 
&- \sum_{i=1}^{m} \frac{g_i^2}{i\Delta_i - \kappa_i / 2} \sigma^{-} \nonumber \\
&- \sum_{i \neq j}^m \frac{g_i g_j J_{ij} e^{i(\phi_i - \phi_j + \theta_{ij})}}{(i\Delta_i - \kappa_i / 2)(i\Delta_j - \kappa_j / 2)} \sigma^{-}.
\label{eq:qubit_dynamics}
\end{align}

The total complex-valued Purcell decay amplitude, \( \Gamma_P \), is given by
\begin{equation}
\Gamma_P = 
\sum_{i=1}^{m} \frac{g_i^2}{i\Delta_i - \kappa_i / 2}
+ \sum_{i < j}^m \frac{2g_i g_j J_{ij} \cos(\phi_i - \phi_j + \theta_{ij})}{(i\Delta_i - \kappa_i / 2)(i\Delta_j - \kappa_j / 2)},
\label{eq:purcell_decay}
\end{equation}
where we have used the fact that $\theta_{ij} = - \theta_{ji}$ to re-index the summation and extract the cosine. Here, \( \Gamma_P \) represents the coherent sum of decay amplitudes from the qubit to the external environment through multiple coupled modes, and corresponds to the complex amplitude appearing in \eqref{eq:qubit_dynamics} for the effective coupling to the environment. The real part corresponds to the actual energy dissipation rate of the qubit into the environment through the resonator modes, while any imaginary component would represent a frequency shift rather than physical decay. The physical Purcell decay rate $\Gamma_{\text{eff}}$ is given by \(2\,\text{Re}(\Gamma_P)\). This factor of 2 accounts for the fact that \(\Gamma_P\) governs the exponential decay of the qubit amplitude \(\sigma^{-}(t) \sim e^{-\Gamma_P t}\), whereas the decay of the excited-state population is determined by the modulus squared, \(|\sigma^{-}(t)|^2 \sim e^{-2\,\text{Re}(\Gamma_P) t}\). Thus
\begin{multline}
\Gamma_{\text{eff}} = 
\underbrace{
\sum_i^m 
\frac{\kappa_i g_i^2}{\Delta_i^2 + \mleft( \frac{\kappa_i}{2} \mright)^2}
}_{\text{Direct decay}} \\
+ \underbrace{
\sum_{i < j}^m 
\frac{
2(\kappa_i \Delta_j + \kappa_j \Delta_i) g_i g_j J_{ij} \cos(\phi_i - \phi_j + \theta_{ij})
}{
(\Delta_i^2 + \mleft( \frac{\kappa_i}{2} \mright)^2)
(\Delta_j^2 + \mleft( \frac{\kappa_j}{2} \mright)^2)
}}_{\text{Interference term}}.
\label{eq:purcell_decay_corrected}
\end{multline}
The first sum describes direct emission into each damped mode, while the second sum accounts for the cross-interference terms. For complete cancellation of radiative decay (i.e., \( \Gamma_{\text{eff}} = 0 \)),  the interference term must exactly cancel the direct term, which occurs at specific qubit frequencies. 

\begin{figure}
\centering
\includegraphics[width=\linewidth]{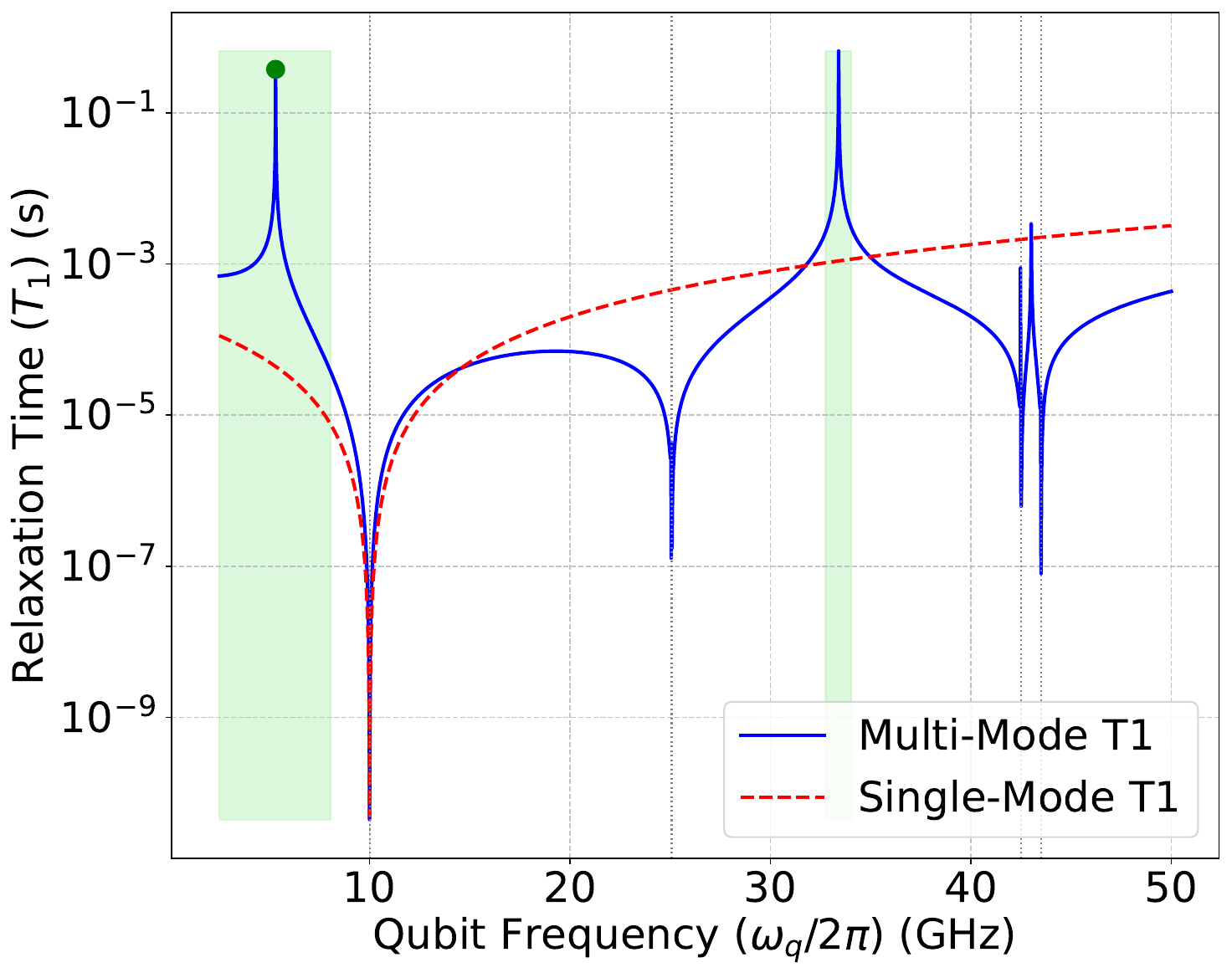}
\caption{Comparison of single-mode and multi-mode Purcell decay rates. Qubit relaxation time \( T_1 \) as a function of qubit frequency \( \omega_q \), showing the single-mode model (red dashed curve) and the full multi-mode interference model (blue solid curve). The qubit is dispersively coupled to a readout resonator at \( \omega_r / 2\pi = \SI{10}{\giga\hertz} \) with linewidth \( \kappa / 2\pi = \SI{8}{\mega\hertz} \). In the multi-mode case, it also interacts with four high-frequency modes at \( \omega_r / 2\pi = 25.04 \), \( 25.08 \), \( 43.52 \), and \( \SI{43.54}{\giga\hertz} \). The qubit-mode coupling strengths are \( g_i / 2\pi = 250,\, 200,\, 180,\, 200 \), and \( \SI{180}{\kilo\hertz} \), and the inter-mode coupling \( J / 2\pi = \SI{180}{\mega\hertz}\). Regions of enhanced lifetime are marked by green bands.}
\label{fig:purd}
\end{figure}

Figure~\ref{fig:purd} presents a comparison between single-mode and multi-mode Purcell decay rates as a function of qubit frequency and inter-mode coupling. The red dashed curve illustrates the single-mode Purcell decay scenario, where the qubit is dispersively coupled to a single readout resonator at \SI{10}{\giga\hertz} with a linewidth of $\kappa = \SI{8}{\mega\hertz}$. As expected, the decay rate increases sharply as the qubit frequency approaches the resonator, following the characteristic inverse-square dependence on detuning.  In contrast, the blue solid curve depicts the multi-mode case, where the qubit is positioned at \SI{6}{\giga\hertz} and additionally interacts with four higher-order resonator modes at 25.04, 25.08, 43.54, and \SI{43.58}{\giga\hertz}. These extra coupling pathways introduce interference between direct and indirect decay processes. Remarkably, this leads to strong suppression of the Purcell decay rate at specific frequencies where destructive interference occurs, corresponding to transmission zeros in the effective decay amplitude. This mechanism operates analogously to a notch filter and allows for passive protection of the qubit from radiative loss. The results confirm that by engineering interference between multiple decay paths—without adding separate Purcell filters—it is possible to significantly enhance the qubit relaxation time $T_1$ across a broad frequency range. This approach enables a hardware-efficient strategy for noise mitigation in superconducting quantum processors. We cross-validate this semi-classical analysis using both an analytical quantum derivation and a density-matrix approach, with all algebraic steps and recursive solutions detailed in Appendices~\ref{app:density_matrix} and~\ref{analytical}.

\section{Purcell decay with microwave drive in multi-mode systems}
\label{sec:AnalysisDrivenPurcell}
We now analyze how a coherent microwave drive modifies the qubit’s radiative decay~\cite{ref42} in the presence of multi-mode interference. Working in a rotating frame at the readout drive frequency $\omega_p$, we assume a single-tone classical coherent drive with amplitude $\epsilon_i$ on each resonator mode $a_i$. We define \( M \) as the undriven Hamiltonian only involving resonator modes [the second and fifth terms in \eqref{eq:multi_mode_H} in the rotating frame], and let \( U \) be the matrix that diagonalizes \( M \). This defines the hybridized modes \( \tilde{a}_k = \sum_i U_{ik} a_i \) with corresponding eigenfrequencies \( \tilde{\Delta}_k \). The full Hamiltonian in the normal-mode basis is
\begin{multline}
\label{eq:Hrot_final}
H_{\text{rot}} = \Delta_{q,d}\, b^\dagger b - \frac{\alpha}{2} b^\dagger b^\dagger b b + \sum_k \tilde{\Delta}_{k,d}\, \tilde{a}_k^\dagger \tilde{a}_k 
\\+ \sum_k \tilde{g}_k \mleft( e^{i \tilde{\phi}_k} b^\dagger \tilde{a}_k + e^{-i \tilde{\phi}_k} b \tilde{a}_k^\dagger \mright)
+ \sum_k \tilde{\epsilon}_k \mleft( \tilde{a}_k + \tilde{a}_k^\dagger \mright).
\end{multline}
Here, $\alpha$ represents the transmon anharmonicity, $\Delta_{q,d} = \omega_q - \omega_p$ is the qubit detuning from the drive frequency, and $\tilde{\Delta}_{k,d} = \tilde{\omega}_k - \omega_p$ are the normal mode detunings from the drive frequency. The effective complex coupling \(\tilde{g}_k e^{i\tilde{\phi}_k} = \sum_i g_i e^{i\phi_i} U^*_{ki}\) preserves all phase information from the original qubit-mode couplings \(g_i e^{i\phi_i}\). The phase and interference effects arising from both the direct couplings and the inter-mode interactions \( J_{ij} \) are implicitly retained through the eigenstructure of \( M \).

To analyze the effect on the qubit from driving the resonator modes, we apply the dispersive approximation to the normal mode Hamiltonian. We assume the dispersive regime where $|\tilde{g}_k| \ll |\tilde{\Delta}_{k,q}|$, where $\tilde{\Delta}_{k,q} = \omega_q - \tilde{\omega}_k$ represents the detuning between the qubit and normal mode $k$,
allowing us to perform a unitary transformation that eliminates the qubit-resonator coupling to first order in $\tilde{g}_k/\tilde{\Delta}_k$. We apply the Schrieffer-Wolff transformation $U_{\text{SW}} = \exp(S)$ with
\begin{equation}
S = \sum_k \mleft(\frac{\tilde{g}_k e^{-i\tilde{\phi}_k}}{\tilde{\Delta}_k} b^\dagger \tilde{a}_k - \frac{\tilde{g}_k e^{i\tilde{\phi}_k}}{\tilde{\Delta}_k} b \tilde{a}_k^\dagger\mright) 
\label{eq:SW_generator}
\end{equation}
This generator eliminates first-order qubit-mode coupling. Higher-order terms involving virtual transitions through anharmonic levels ($\ket{2}$, $\ket{3}$, etc.) contribute to the complete dispersive transformation and are essential for deriving the anharmonicity-dependent shifts $\chi_k$ in Eq.~(\ref{eq:chi_k}). The complete derivation including these anharmonic corrections is detailed in standard cQED references~\cite{ref4,ref38}. Applying this transformation to second order and accounting for the transmon anharmonicity $\delta_q$, we obtain
\begin{multline}
\label{eq:H_disp}
H_{\text{disp}} = 
\mleft( \Delta_q + \sum_k \chi_k\, \tilde{a}_k^\dagger \tilde{a}_k \mright) b^\dagger b 
- \frac{1}{2} \alpha \, b^\dagger b^\dagger b b \\
+ \sum_k \tilde{\Delta}_k\, \tilde{a}_k^\dagger \tilde{a}_k 
+ \sum_k \tilde{\epsilon}_k \mleft( \tilde{a}_k + \tilde{a}_k^\dagger \mright) 
+ \sum_k \chi_k' \mleft( \tilde{a}_k^\dagger \tilde{a}_k \mright)^2 \\ 
+ \sum_{k<l} \chi_{kl}\, \tilde{a}_k^\dagger \tilde{a}_k\, \tilde{a}_l^\dagger \tilde{a}_l 
+ \mathcal{O}\mleft( \frac{\tilde{g}_k^3}{\tilde{\Delta}_k^2} \mright),
\end{multline}
where the dispersive shifts are given by~\cite{ref38}
\begin{equation}
\chi_k = -\frac{\tilde{g}_k^2}{\tilde{\Delta}_k} \mleft(1 + \frac{\alpha}{\tilde{\Delta}_k - \alpha}\mright) = -\frac{\tilde{g}_k^2\alpha}{\tilde{\Delta}_k(\tilde{\Delta}_k - \alpha)}.
\label{eq:chi_k}
\end{equation}
The anharmonicity term has been absorbed into the definition of the dispersive shifts~$\chi_k$ and higher-order corrections through the Schrieffer-Wolff transformation. In the dispersive limit $|\tilde{g}_k| \ll |\tilde{\Delta}_k|$, $|\alpha| \ll |\tilde{\Delta}_k|$, the renormalization of the bare anharmonicity~$\alpha$ is negligible compared to the dispersive effects. For analyses strictly within the single-excitation subspace, the effects of these higher levels are captured in the modified dispersive shift expressions without the need to explicitly include states beyond \(|0\rangle\) and \(|1\rangle\). The derivation of the cross-Kerr interaction is detailed in Appendix~\ref{crossker} for the simplified two-level case.

Under coherent driving, each normal mode reaches a steady state with a mean photon number. We solve for this steady state by moving to a displaced frame defined by the displacement operators
\begin{equation}
D_k(\alpha_k) = \exp(\alpha_k\tilde{a}^\dagger_k - \alpha^*_k\tilde{a}_k),
\end{equation}
where \( \alpha_k \) are the complex amplitudes of the coherent states in each mode. In the presence of dissipation with rates \( \tilde{\kappa}_k \) for each normal mode, the steady-state amplitudes are given by
\begin{equation}
\alpha_k = \frac{\tilde{\epsilon}_k}{i\tilde{\Delta}_k + \tilde{\kappa}_k/2},
\label{photon_linear}
\end{equation}
with corresponding mean photon numbers \( \bar{n}_k = |\alpha_k|^2 \). The total mean photon number is 
\begin{equation}
\bar{n} = \sum_k \bar{n}_k.
\end{equation}

In the dispersive limit, the shifted qubit frequency induced by a populated mode \( k \), where $\bar{n}_k = \langle \tilde{a}_k^\dagger \tilde{a}_k \rangle$, is
\begin{equation}
\omega_{q,\text{eff}}(\{\bar{n}_k\}) \approx \omega_q + \sum_k 2\chi_k \bar{n}_k,
\end{equation}
where we have defined an effective per-photon shift \( \chi_{\text{eff}} = \frac{\sum_k \chi_k \bar{n}_k}{\bar{n}} \) with total mean photon number $\bar{n} = \sum_k \bar{n}_k$, which depends on the distribution of photons across normal modes.
For simplicity, when a single mode dominates (typically the readout resonator), \( \chi_{\text{eff}} \approx \chi_r \). 

The Purcell decay rate of the qubit arises from its coupling to the normal modes, which in turn are coupled to the external environment.
The zero-drive radiative decay rate, which can be obtained either via the linearized Heisenberg--Langevin equation approach in Section~\ref{sec:AnalysisSingleExc}, or equivalently via the perturbative expansion in the normal-mode basis detailed in Appendix~\ref{app:density_matrix}, is given by
\begin{multline}
\Gamma_{\text{Purcell}}(0) = 
\sum\limits_i \frac{\kappa_i g_i^2}{\Delta_i^2 + \mleft( \frac{\kappa_i}{2} \mright)^2} \\
+ \sum\limits_{i < j} \frac{\mleft( \kappa_i \Delta_j + \kappa_j \Delta_i \mright) g_i g_j J_{ij} 
\cos(\phi_i - \phi_j + \theta_{ij})}
{\mleft[ \Delta_i^2 + \mleft( \frac{\kappa_i}{2} \mright)^2 \mright] \mleft[ \Delta_j^2 + \mleft( \frac{\kappa_j}{2} \mright)^2 \mright]}.
\label{eq:Gamma_zero}
\end{multline}
We normalize the driven decay rate by its zero-drive value
\begin{equation}
\frac{\Gamma_{\text{Purcell}}(\bar{n}_k)}{\Gamma_{\text{Purcell}}(0)} = \frac{\text{Decay rate with drive}}{\text{Decay rate without drive}}.
\label{eq:purcell_decay_normalized}
\end{equation}
This normalization eliminates common prefactors and isolates the functional dependence on the drive strength. 

To understand how the Purcell rate scales with drive strength in our multi-mode system, we consider the scenario where the qubit is far detuned from all modes (\( \tilde{\Delta}_k \gg \tilde{\kappa}_k \)), and the drive primarily populates one dominant mode (typically the readout resonator). For simplicity, we define the effective dispersive shift as \( \chi = \chi_{\text{eff}} \). In this regime, the direct decay term, assuming AC Stark shift \( 2\chi_{\text{eff}}\bar{n} \), is
\begin{multline}
\frac{\tilde{\kappa}_k|\tilde{g}_k|^2}{(\tilde{\Delta}_k + 2\chi_{\text{eff}}\bar{n})^2 + (\tilde{\kappa}_k/2)^2}
\approx
\frac{\tilde{\kappa}_k|\tilde{g}_k|^2}{(\tilde{\Delta}_k + 2\chi_{\text{eff}}\bar{n})^2}
\\
\propto
\frac{1}{\left[1 + O(\bar{n}/n_{\text{crit}})\right]^2}
\label{eq:direct_decay_scaling}
\end{multline}
For the interference term, when the frequency shifts are small compared to the detunings, we have
\begin{multline}
\frac{\mleft(\tilde{\kappa}_k \tilde{\Delta}_l + \tilde{\kappa}_l \tilde{\Delta}_k \mright) 
|\tilde{g}_k||\tilde{g}_l| |J_{ij}| 
\cos[\arg(\tilde{g}_k) - \arg(\tilde{g}_l)]}{
\mleft[\tilde{\Delta}_k^2 + (\tilde{\kappa}_k/2)^2 \mright]\mleft[\tilde{\Delta}_l^2 + (\tilde{\kappa}_l/2)^2\mright]}
\\
\propto
\frac{1 + \frac{2\chi_{\text{eff}}\bar{n}(\tilde{\kappa}_k + \tilde{\kappa}_l)}{\tilde{\kappa}_k \tilde{\Delta}_l + \tilde{\kappa}_l \tilde{\Delta}_k}}{\mleft(1+\frac{2\chi_{\text{eff}}\bar{n}}{\tilde{\Delta}_k}\mright)^2\mleft(1+\frac{2\chi_{\text{eff}}\bar{n}}{\tilde{\Delta}_l}\mright)^2}
\label{eq:interference_term_scaling}
\end{multline}
In the limit where \( \bar{n} \ll \frac{\tilde{\Delta}_k}{2\chi_{\text{eff}}} \) but not negligible, this yields
\begin{equation}
\propto
\frac{1}{\mleft[1 + O(\bar{n}/n_{\text{crit}})\mright]^{3-4}},
\label{eq:interference_term_approx}
\end{equation}
where \( n_{\text{crit}} = \frac{|\tilde{\Delta}_k|}{2|\chi_{\text{eff}}(0)|} \). This indicates that the interference term will generally scale more steeply with drive power than the direct term.

However, the actual scaling will depend on the specific parameters of the system, including the relative magnitudes of detunings, coupling strengths, and the precise engineered interference conditions. When destructive interference is present at zero drive, the combined scaling could be approximately
\begin{equation}
\frac{\Gamma_{\text{Purcell}}(\bar{n})}{\Gamma_{\text{Purcell}}(0)} \sim \mleft(1 + \frac{\bar{n}}{n_{\text{crit}}}\mright)^{-\alpha},
\label{eq:combined_scaling}
\end{equation}
where $\alpha$ is a positive number that depends on the relative contributions of direct and interference terms and the specific parameters of the engineered interference condition, providing enhanced protection against Purcell decay under driving.

\begin{figure}
\includegraphics[width=\linewidth]{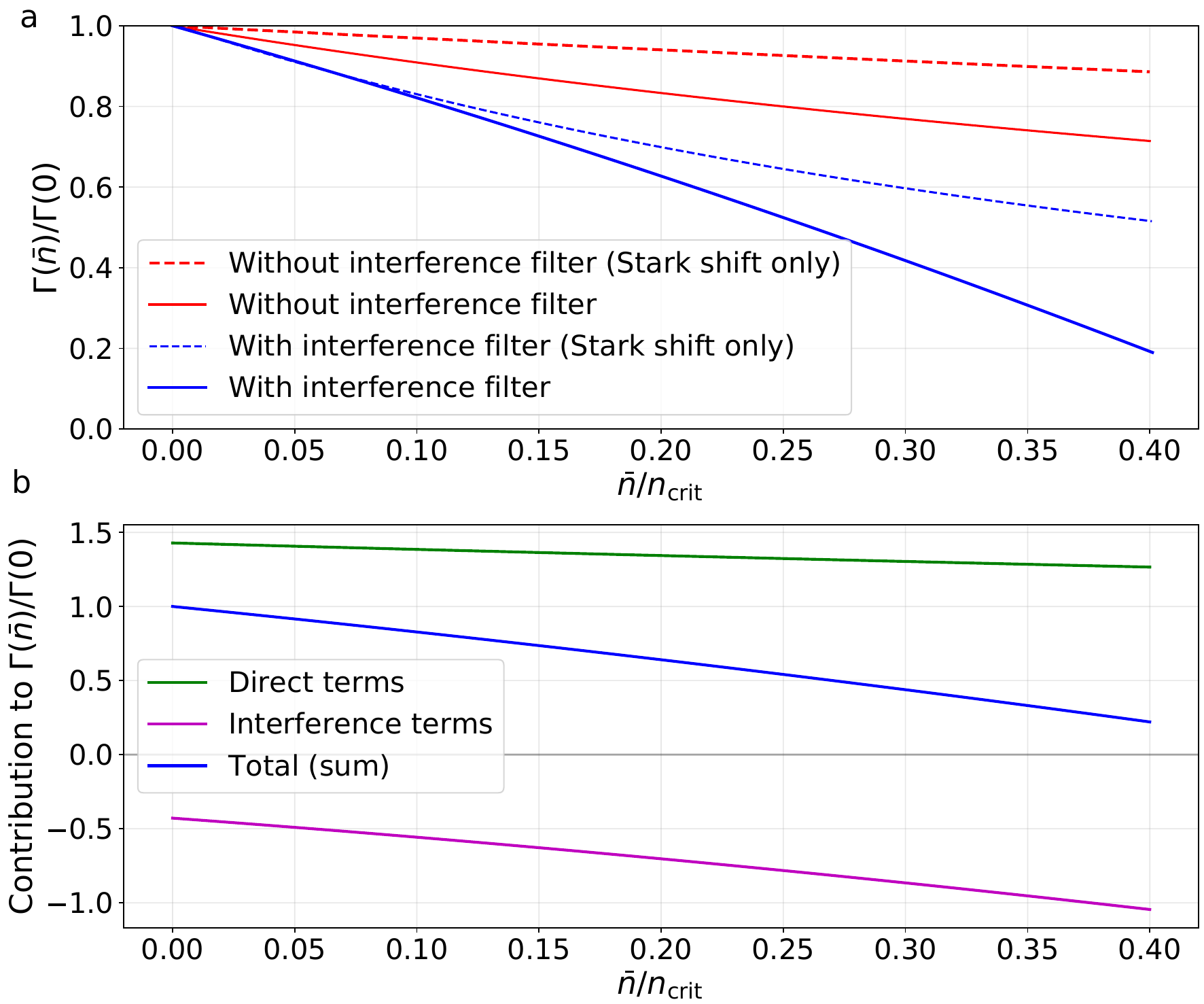}
\caption{\label{n_purcell_decay_rate} Purcell-decay suppression in multi-mode interference systems under drive. (a) Numerical and analytical comparison of the normalized Purcell decay rate $\Gamma(\bar{n})/\Gamma(0)$ as a function of dimensionless drive strength $\bar{n}/|\Delta/(2\chi_{\text{eff}}(0))|$. Solid curves show full numerical results, while dashed lines show analytical predictions considering only AC-Stark shift effects. The blue curves correspond to the multi-mode case with engineered mode-mode coupling $J_{ij}$, designed to achieve destructive interference between decay pathways. The red curves correspond to the single-mode case without interference. A steeper suppression of the Purcell rate is observed in the multi-mode system, scaling approximately as $(1 + \bar{n}/n_{\text{crit}})^{-2}$, compared to the $(1 + \bar{n}/n_{\text{crit}})^{-1}$ scaling in the single-mode case. (b) Decomposition of the multi-mode Purcell decay rate into its constituent parts. The green line represents the direct decay terms that decrease with drive strength. The magenta line shows the interference terms, which become increasingly negative with higher dimensionless drive strength, enhancing the suppression effect. The blue line shows the total normalized Purcell rate, which approaches zero at higher drive powers due to the cancellation between direct and interference terms. The simulations were performed using the following representative dimensionless parameters normalized by the qubit-resonator detuning scale $\Delta$: qubit--resonator coupling strength \( g = 0.25 \), resonator decay rate \( \kappa = 0.02 \), AC Stark shift coefficient \( \chi = 0.03 \), and critical photon number \( n_{\mathrm{crit}} = 2.0 \). The simulation explicitly models a multi-mode system with the primary readout resonator and two higher-frequency electromagnetic modes, with inter-mode couplings $J_{ij} = 0.05$ engineered to achieve destructive interference between decay pathways.}
\end{figure}

The numerical and analytical results presented in \figref{n_purcell_decay_rate} illustrate the drive-dependent suppression of the qubit Purcell decay rate in the presence of multi-mode interference. We simulate the non-unitary dynamics of the qubit-mode system using the effective non-Hermitian Hamiltonian \( H_{\mathrm{eff}} \) derived in Appendix~\ref{app:density_matrix}, \eqref{eq:H_eff_lab_fig}. We restrict the Hilbert space to the single-excitation manifold \(\{ \ket{e,0}, \ket{g,1_1}, \ket{g,1_2}, \dots, \ket{g,1_m} \}\), where the qubit is either excited with all modes in vacuum, or the qubit is in the ground state with one photon occupying mode \( i \). The anharmonic effects of the transmon are incorporated through the anharmonicity-dependent dispersive shifts $\chi_k$ given by Eq.~(\ref{eq:chi_k}), which capture the virtual transitions through higher qubit levels $\ket{2}$, $\ket{3}$, etc. This effective single-excitation treatment is valid in the dispersive regime, where direct population of higher qubit levels is suppressed. The Hamiltonian includes the qubit-mode coupling terms \( g_i e^{i\phi_i} \), complex inter-mode hopping terms \( J_{ij} e^{i\theta_{ij}} \), and resonator dissipation modeled via the imaginary potentials \( -i\kappa_i/2 \). We numerically construct this \( (1 + m) \times (1 + m) \) matrix and compute its full complex eigenspectrum. 
To include the effect of the coherent drive, we first compute the steady-state intracavity photon populations \( \bar{n}_i \) in each normal mode using the linear-response expression \( \bar{n}_i = \mleft| \frac{\tilde{\epsilon}_i}{i\tilde{\Delta}_i + \tilde{\kappa}_i/2} \mright|^2 \), as described in \eqref{photon_linear}. These photon populations induce an AC Stark shift of the qubit frequency, modifying the detunings as $\tilde{\Delta}_{i,q} \to \tilde{\Delta}_{i,q} + 2\chi_{\text{eff}} \bar{n}$. The Stark-shifted detunings are inserted directly into the Hamiltonian \( H_{\mathrm{eff}} \), which is then diagonalized to extract the complex eigenvalue \( \lambda_e \) continuously connected to the bare qubit state in the limit \( g_i \to 0 \). The qubit decay rate is extracted as \( \Gamma(\bar{n}) = -2\, \mathrm{Im}(\lambda_e) \). We repeat this procedure for a sweep of drive powers by varying \( \bar{n} \), updating the detunings accordingly, and re-diagonalizing the Hamiltonian at each step. This self-consistent approach captures both AC Stark shift and multi-mode interference effects beyond the lowest-order perturbative treatment within the single-excitation subspace.
In contrast, the analytical results are obtained using the third-order perturbative formula from Appendix~\ref{app:density_matrix}, \eqref{qc:purcell_decay_corrected}, with the same Stark-shifted detunings inserted. The agreement and deviation between the two approaches are shown in \figref{n_purcell_decay_rate}, highlighting the nonlinear response and the critical role of interference terms.

The total decay rate, given in \eqref{eq:purcell_decay_normalized}, comprises two components: the direct decay terms and the interference terms. Figure~\ref{n_purcell_decay_rate} confirms our theoretical  prediction: the blue curves (with engineered interference) show significantly steeper suppression compared to the red curves (without interference), consistent with the scaling laws derived in \eqref{eq:interference_term_approx}. The discrepancy between the numerical results (solid lines) and analytic predictions (dashed lines) highlights the essential role of interference terms, which are absent in Stark-only models. The decomposition shown in \figpanel{n_purcell_decay_rate}{b} further confirms that the suppression arises predominantly from the interference pathway, whose contribution becomes increasingly negative with drive. These theoretical results suggest that in an ideal system, the AC Stark shift could potentially not only suppress the direct decay channel but also amplify engineered destructive interference. However, practical implementations would need to account for realistic drive conditions, where higher modes may experience significantly different drive amplitudes due to circuit filtering and drive-line transmission-frequency characteristics.

\section{Finite-element simulation of an intrinsically protected superconducting qubit device}
\label{sec:Simulation}

\begin{figure}
      \includegraphics[width=\linewidth]{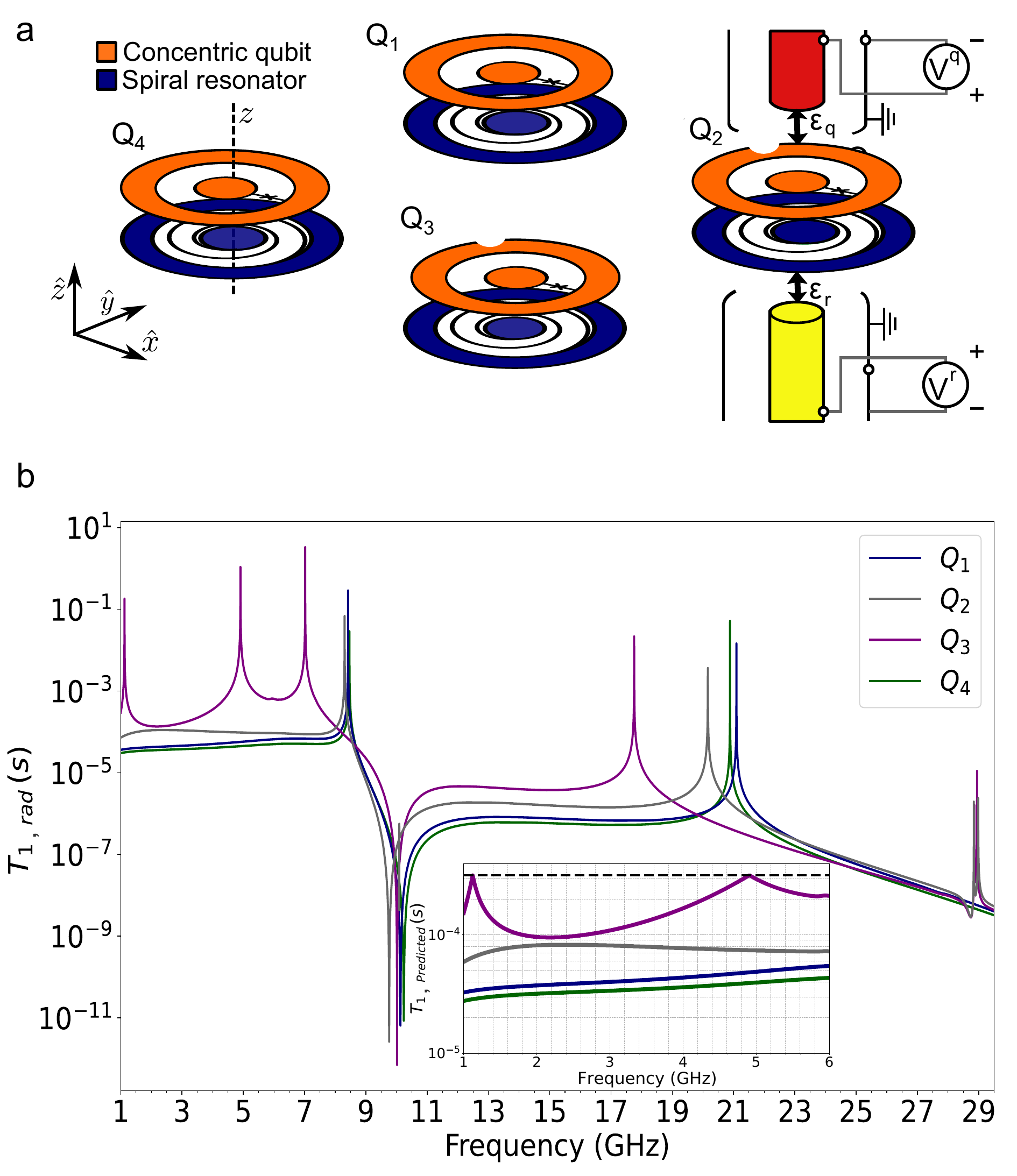}
        \caption{Four-qubit superconducting device with engineered 'mouse bite' features. (a) Circuit layout illustration (not to scale) showing two standard transmons ($Q_1$ and $Q_4$) and two modified transmons ($Q_2$ and $Q_3$) with semi-circular cuts to break symmetry and activate multi-path interference. Out-of-plane wiring is shown for $Q_2$. (b) Simulation results of the fabricated and measured device using measured $\kappa$ values (see Table~\ref{4Q-MB_Reslts}) showing the predicted total relaxation time $T_{1,\text{Predicted}} = \left( \frac{1}{T_{1,\text{Intrinsic}}} + \frac{1}{T_{1,\text{rad}}} \right)^{-1}$, where $T_{1,\text{Intrinsic}} = Q_{\text{internal}}/\omega_q$ with an estimated qubit internal quality factor $Q_{\text{internal}} = 10^7$. The inset shows an enlarged view of the individual qubit $T_{1,\text{Predicted}}$ values, with $Q_3$ exhibiting a real-frequency transmission zero (notch) at 4.9 GHz and a complex transmission zero (notch) at 2 GHz. $Q_3$ exhibits a rich $T_1$ structure with multiple peaks and dips as a function of frequency. These features originate from frequency-dependent interference  effects shaped by the local electromagnetic environment of the qubit. In particular, the spatial separation between  the drive line, the qubit capacitor pads, and the coupled readout resonator places $Q_3$ in an optimal regime to support  engineered interference.
        }
        \label{FB}
\end{figure}

We now move on to perform finite-element (FE) simulations of the actual fabricated device model of a four-qubit superconducting circuit device, shown in the inset of \figref{FB}. The simulations were performed using the eigenmode and frequency-domain solvers in Ansys HFSS. The model is based on an architecture in which universal quantum control and readout have been demonstrated~\cite{ref35, ref48, ref49}. The model consists of a $2 \times 2$ array of coaxial transmon qubit islands on a silicon substrate, enclosed by a perfectly conducting cavity. The distance between the inner islands of adjacent qubits is \SI{1.5}{\milli\metre}, and two of the qubit islands are perturbed with a `mouse bite' feature to break geometrical symmetry and activate multiple-path cancellation.
Each qubit island is capacitively coupled to an off-chip coaxial drive line. The qubits have JJ capacitance $C_J \approx \SI{100}{\femto\farad}$ and JJ inductance $L_J \approx \SI{10}{\nano\henry}$ at fundamental transition frequency $f_{01} = \SI{5}{\giga\hertz}$, corresponding to a Josephson energy to charging energy ratio $E_J / E_C \approx 80$, with designed readout resonator frequencies around $f_r = \SI{9}{\giga\hertz}$.

In the absence of a `mouse bite' perturbation, the fundamental mode of the transmon qubit couples weakly to the geometric modes of the system, and the coupling \( J_{ij} \) between the higher geometric modes is effectively zero. However, in the presence of the `mouse bite' feature, the effective coupling between the transmon fundamental mode and the higher modes, as well as the coupling between the higher modes themselves, increases. The relative phase condition is controlled by carefully positioning the 'mouse bite' feature relative to the orientation of the Josephson junction. For degenerate modes of the qubit capacitive pads, the coupling \( J_{ij} \) between  is maximized at 45$^\circ$, 135$^\circ$, 225$^\circ$, and 315$^\circ$ relative to the junction, while the coupling sign (inductive or capacitive) is determined by whether the feature is an extrusion (outward protrusion) or bite (inward cut) and its specific angular position. The coupling strength \( J_{ij} \) is controlled by varying the size and position of the `mouse bite' feature (increasing the size will generally increase $J_{ij}$), the capacitive pad geometries including the width and radius of the outer ring, and the relative positioning of these elements with respect to the Josephson junction. This control mechanism is specific to the transmon layout, as the higher-frequency electromagnetic modes depend critically on the transmon's geometric shape, dimensions, and symmetry properties. For further details, see Appendix~\ref{app:em}.

\begin{table*}
\centering
\small  
\caption{Experimentally measured device parameters. 
$Q_{\text{int}}$ is the internal quality factor of the readout resonator; $\kappa_{\text{ext}}$ is the external coupling rate extracted from resonator transmission measurements; $\chi$ is the qubit-resonator dispersive shift; $\alpha$ is the qubit anharmonicity (defined as $\omega_{12} - \omega_{01}$); $T_1$, $T_{2R}$, and $T_{2E}$ are energy relaxation, Ramsey coherence, and Hahn echo coherence times, respectively. Parentheses indicate the fit uncertainty from exponential decay fits. The the standard deviation of $T_1$, $T_{2R}$, and $T_{2E}$ extracted from multiple repeated measurements is shown in parentheses after coherence-time values, capturing statistical fluctuations over time. $\mathcal{F}$ is the single-qubit gate fidelity measured via randomized benchmarking on XY Clifford gates.}
\vspace{0.2em}
\resizebox{1\textwidth}{!}{%
\begin{tabular}{lccccccccccc}
\toprule
& \( \omega_r/2\pi~[\mathrm{GHz}] \) 
& \( \omega_q/2\pi~[\mathrm{GHz}] \) 
& \( Q_{\text{int}}~[10^4] \) & \( \kappa/2\pi~[\mathrm{MHz}] \)
& \( \kappa_{\text{ext}}/2\pi~[\mathrm{MHz}] \) 
& \( \chi/2\pi~[\mathrm{kHz}] \) 
& \( \alpha/2\pi~[\mathrm{MHz}] \) 
& \( T_1~[\mu\mathrm{s}] \) 
& \( T_{2R}~[\mu\mathrm{s}] \) 
& \( T_{2E}~[\mu\mathrm{s}] \) 
& \( \mathcal{F}~[\%] \)  \\
\midrule
\( Q_1 \) & 8.81 & 5.01 & 2.11 & 7.12 & 6.78 & -348 & -231 & 34\,(3) & 26\,(2) & 45\,(6) & 99.72  \\
\( Q_2 \) & 8.90 & 4.84 & 2.63 & 5.41 & 5.09 & -352 & -232 & 34\,(7) & 25\,(4) & 41\,(10) & 99.72  \\
\( Q_3 \) & 9.01 & 5.24 & 2.76 & 0.51 & 0.18 & -329 & -228 & 66\,(30) & 37\,(7) & 88\,(21) & 99.82 \\
\( Q_4 \) & 9.15 & 4.95 & 2.81 & 2.13 & 1.82 & -247 & -229 & 21\,(4) & 9\,(3) & 18\,(4) & 99.71 \\
\bottomrule
\end{tabular}}
\label{4Q-MB_Reslts}
\end{table*}

The number and position of the transmission zeros in the spectrum, i.e., the peaks in \figref{FB}, are determined by the detuning between the qubit and resonator modes, the coupling strengths, the relative phases of the interacting modes, and the external coupling rate \(\kappa_{\text{ext}}\), which acts as a critical control parameter. The influence of these parameters in turn depends on factors like the transmon geometry, making it hard to provide general design guidelines. In our implementation, the external coupling rates are extracted from experiment and found to be \(\kappa/2\pi = \{7.12, 5.41, 0.51, 2.13\}~\text{MHz}\) for qubits \(Q_1\) through \(Q_4\), respectively. By tuning \(\kappa\), we modulate the bandwidth and strength of interaction with the environment, thereby directly influencing the location of the transmission zeros. In this implementation, increasing $\kappa$ shifts the transmission zeros from the complex domain to the real-frequency axis, and eventually the peaks split and move away from each other, requiring numerical optimization. Alternative tuning knobs include the physical layout of the capacitor pads, the qubit–resonator spacing, and the symmetry-breaking notch geometry. In this implementation, the mouse-bite feature has a radius of \SI{0.11}{\milli\metre}, positioned \(135^\circ\) from the Josephson junction, and plays a key role in shaping interference effects across the frequency spectrum.

This modification introduces a notable perturbation in the capacitive network between the ground plane, the control lines, and the qubit and resonator pads. Notably, we observe a two-order-of-magnitude increase in the coupling between the control lines, as well as between the qubit control line and the resonator, indicating a substantial enhancement in the strength of cross-coupling interactions. The modified capacitance matrix is summarized in \eqref{eq:deltaC_matrix}. We then analyze the admittance response of both the original and modified devices using a driven-terminal simulation, where an excitation port is defined at the qubit junction to probe its microwave environment. The results, illustrated in \figref{FB}, reveal that Q3 exhibits two cancellation zeros at \SI{5.83}{\giga\hertz} and \SI{6.3}{\giga\hertz} in the modified device, in contrast to the absence of cancellation zeros in the original configuration. Additionally, higher-frequency modes, whose couplings to the lower-frequency modes have been activated, emerge across the simulated broadband frequency spectrum, demonstrating multi-path interference enabled by the symmetry-breaking geometry.

\section{Experiment}
\label{sec:Experiment}

\begin{figure}
\includegraphics[width=\linewidth]{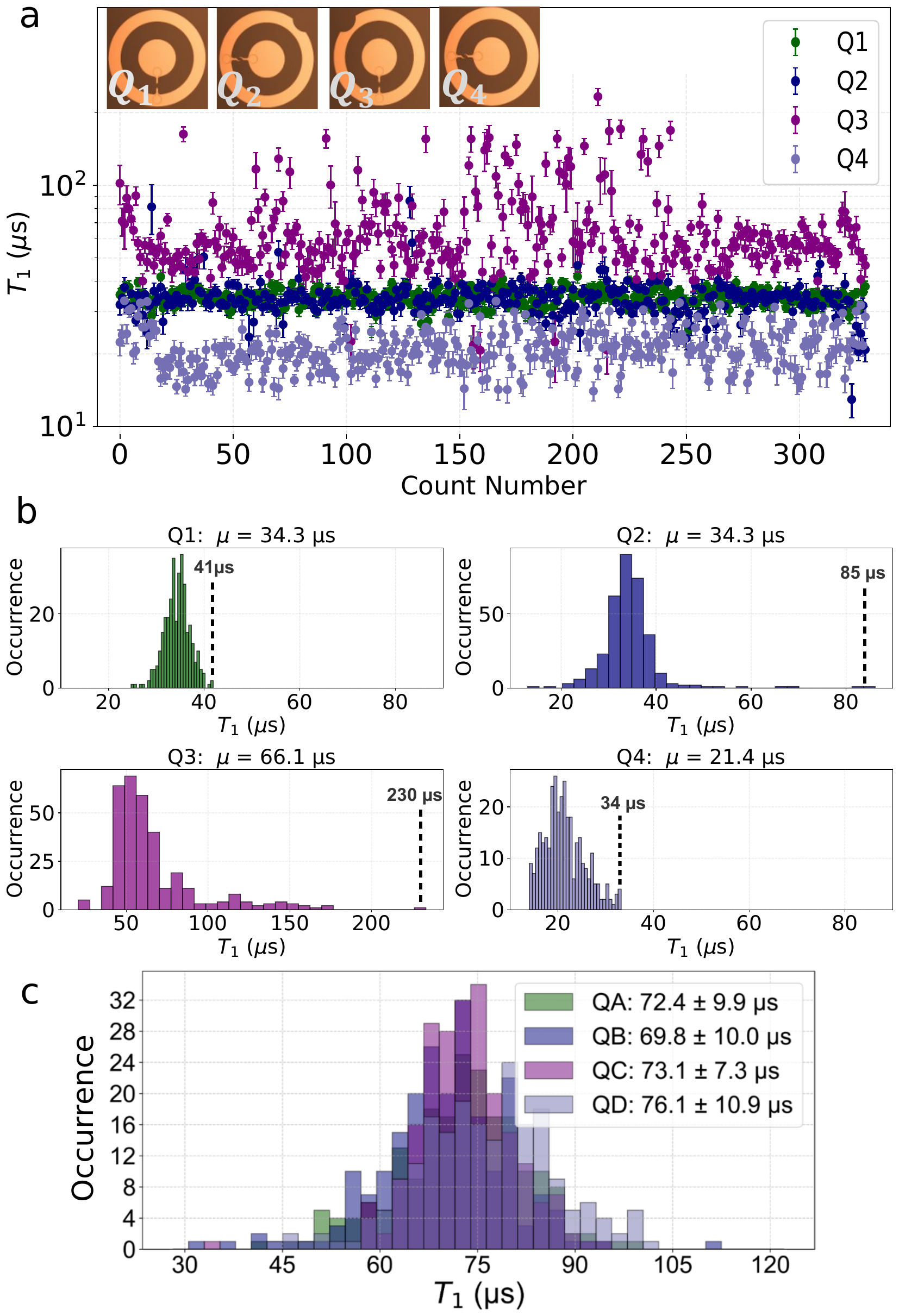}
\caption{\label{fig:epsart} Energy relaxation time \( T_1 \) statistics for two four-qubit superconducting devices. (a) Recorded \( T_1 \) measured over 12 hours for four qubits (Q1–Q4) on the first device. Inset: optical images of the four qubits. (b) Histograms of the same \( T_1 \) data set for each qubit, showing variations in both the mean and spread across qubits. (c) Combined histogram of the data set from a similar four-qubit device, with a single-pole 3D re-entrant cavity Purcell filter on the readout side. The qubits are coupled to individual readout resonators with external coupling rates \( \kappa_{\text{ext}} / 2\pi \approx \{0.88,\, 1.97,\, 2.30,\, 1.61\}~\text{MHz} \) and qubit frequencies \( \omega_q / 2\pi \approx \{4.33,\, 4.17,\, 4.27,\, 4.23\}~\text{GHz} \). All other device parameters are similar to those reported in Table~\ref{4Q-MB_Reslts}.
}
\label{meas}
\end{figure}

In this section, we provide experimental evidence for suppression of the Purcell decay rate using the methods developed in the previous sections on a device with four fixed-frequency transmon qubits. The inset of Figure~\figpanelNoPrefix{meas}{a} shows optical images of the transmon qubits with/without the `mouse-bite' feature. The device layout is shown in Figure~\figpanelNoPrefix{FB}{a}, with the enclosure bases features detailed in Figure 1 of~\cite{ref48}. In our setup, control signals are routed to qubits and resonators by UT47-type coaxial cables with characteristic impedance $Z_0 = \SI{50}{\ohm}$. The distances between the inner conductor of the readout coaxial cables and the readout resonators R1--R4 are set to 225, 270, 570, and \SI{320}{\micro\metre}, corresponding to extracted coupling rates of $\kappa/2\pi = 7.1$, 5.4, 0.51, and \SI{2.1}{\mega\hertz}, respectively. The external coupling rates \( \kappa_{\text{ext}} / 2\pi \) are reported in Table~\ref{4Q-MB_Reslts}. The quantities $\omega_{q,n}$, $\alpha_n$, $\omega_{r,i}$, and $\chi_{ni}$ were determined using standard spectroscopic measurements and Ramsey measurements; \( \omega_{q,n} \) and \( \alpha_n \) denote the frequency and anharmonicity of the \( n \)th qubit, \( \omega_{r,i} \) the frequency of the \( i \)th resonator mode, and \( \chi_{ni} \) the dispersive shift between qubit \( n \) and mode \( i \). We note that Q3 exhibits a significantly reduced external coupling rate $\kappa/2\pi = \SI{0.51}{\mega\hertz}$ compared to Q1, Q2 and Q4 (\SI{6.7}{\mega\hertz}, \SI{7.1}{\mega\hertz}, and \SI{2.1}{\mega\hertz}), which contributes to its enhanced $T_1$ performance. However, the simulation results (Figure~\ref{FB} inset) show that the predicted $T_1$ improvement from reduced coupling alone cannot fully account for the observed enhancement, indicating additional contributions from multi-mode interference effects. For example, the predicted $T_1$ for $Q_3$ without additional interference effect is estimated to be $\SI{90}{\micro\second}$.

We simultaneously measured the relaxation times $T_1$ of the four qubits repeatedly over a period of 12 hours. The consecutive measured values and the resulting histograms are shown in \figref{fig:epsart}. The characteristic dephasing times $T_{2R}$ and $T_{2E}$ were measured using standard Ramsey and Hahn echo pulse sequences, performed separately on each qubit and repeated over approximately 12 hours per qubit. We perform randomized single-qubit benchmarking (RB) on all four qubits using a combination of \SI{60}{\nano\second} duration (\SI{50}{\nano\second} Blackman envelope~\cite{cao2024} with \SI{10}{\nano\second} buffer) physical gates $I$, and $X_{\pi/2, \pi}$ with derivative removal by adiabatic gate (DRAG) pulse shaping~\cite{ref50} and virtual $Z$ gates. Single-shot readout was performed for all the RB experiments. The RB protocol was run at 31 Clifford sequence lengths and for $k = 80$ different sequences of Clifford gates. Each of the $31 \times 80$ experiments was repeated 5000 times to build statistics. The resulting single-qubit gate fidelities and the rest of the device parameters are summarized in Table~\ref{4Q-MB_Reslts}.

We analyze the experimentally observed $T_1$ values presented in \figref{fig:epsart} to (i) extract the decay rate from other non-radiative loss mechanisms and (ii) examine if the observations are consistent with a suppressed Purcell decay. To this end, we model the total energy relaxation as arising from two statistically independent channels: radiative decay through the Purcell effect, and non-radiative loss mechanisms such as dielectric loss or quasiparticle tunneling. The total decay rate is expressed as
\begin{equation}
\frac{1}{T_1} = \frac{1}{T_{1,\mathrm{Purcell}}} + \frac{1}{T_{1,\mathrm{others}}}.
\label{eq:T1_total}
\end{equation}
Since all qubits are fabricated on a common substrate and measured in a shared cryogenic and electromagnetic environment, we assume \( T_{1,\mathrm{others}} \) to be approximately uniform across the four qubits. We note that this assumption may oversimplify the complexity of non-radiative mechanisms such as two-level fluctuators, which can vary significantly between individual qubits due to local microscopic defects. To estimate the non-radiative ceiling, we examine the qubit with the best performance. From the histogram of Q3 in \figpanel{fig:epsart}{b}, which shows the highest $T_1$ values among the four qubits, we observe an upper bound of the distribution \( T_{1,Q3}^{\mathrm{(max)}} \approx \SI{230}{\micro\second} \), a standard deviation \( \sigma_{Q3} \approx \SI{30}{\micro\second} \), and a mean value of \( T_{1,Q3} \approx \SI{66}{\micro\second} \).

To estimate the non-radiative ceiling, we measure a similar device to the one shown in the inset of \figpanel{fig:epsart}{a}, with a 3D re-entrant cavity and a 4:1 Purcell-filtered multiplexer reported in Ref.~\cite{ref24}. This device does not include intrinsic multiple-path interference—that is, the distances between the readout drive pins and the readout resonators range between \SI{0.35}{\milli\metre} and \SI{0.40}{\milli\metre}, and the diameter of the pins is approximately \SI{1}{\milli\metre}. These parameters are outside the regime for multiple-path Purcell cancellation, which would require a smaller pin diameter $<\SI{0.5}{\milli\metre}$ and pin distance $<\SI{0.425}{\milli\metre}$ for the chosen geometries of the transmon capacitive pads and 'mouse bite' feature and orientation. Crucially, the use of the 4:1 Purcell filter significantly suppresses radiative decay from the readout resonators, thereby allowing non-radiative mechanisms to dominate the observed relaxation times. In contrast, the device in \figpanel{fig:epsart}{a} does not employ external Purcell filters, demonstrating that intrinsic multi-mode interference can potentially replace conventional Purcell filtering while saving chip area. The resulting histograms of measured \(T_{1}\) values are shown in \figpanel{fig:epsart}{c}. Examining the upper bound of the \(T_{1}\) distributions across all four qubits, and noting their narrow spread and consistent shape, we infer that all four qubits are governed by a shared non-radiative ceiling. Based on this, we estimate:
\begin{equation}
    T_{1,\mathrm{others}} \approx \SI{70}{\micro\second}.
    \label{eq:T1_other}
\end{equation}

Furthermore, we examine whether the Purcell-decay suppression seen for qubit Q3 in simulations is consistent with the experimental observations in \figref{fig:epsart}. As shown in the simulation results (Fig.~\ref{FB} inset), Q3 exhibits predicted \(T_{1}\) enhancement peaks, with one prominent peak occurring around \SI{4.9}{\giga\hertz}. Given that Q3 operates at \SI{5.24}{\giga\hertz} (Table~\ref{4Q-MB_Reslts}), it lies in a frequency region where the simulations predict improved coherence due to multi-mode interference effects. First, we draw attention to the fact that the observed average $T_1$ values for qubits Q1, Q2, and Q4 are $T_1 = \SI{34}{\micro\second}$, $\SI{34}{\micro\second}$, and $\SI{21}{\micro\second}$, respectively respectively, which is almost half of that for qubit Q3. Notably, qubit Q2 shows a larger variance than Q1 and Q4 which could be attributed to partial Purcell suppression and is consistent with the simulation data in \figref{FB}, with a maximum recorded \( T_1 \sim \SI{85}{\micro\second} \) compared to \( T_1 \sim \SI{40}{\micro\second} \) for Q1 and Q4. The improved $T_1$ is consistent with a suppressed Purcell decay. Using Eqs.~(\ref{eq:T1_total}) and (\ref{eq:T1_other}), we find that for qubits Q1 and Q4, the estimated Purcell-limited relaxation times are approximately $T_{1,\mathrm{Purcell}} \approx \SI{66}{\micro\second}$ and $\SI{30}{\micro\second}$, respectively. For Q2, we estimate $T_{1,\mathrm{Purcell}} \approx \SI{66}{\micro\second}$. In contrast, for Q3 we estimate $T_{1,\mathrm{Purcell}} \approx \SI{1.2}{\milli\second}$, indicating significantly suppressed Purcell decay with Q3 achieving \SI{94}{\percent} of the estimated intrinsic limit.

Second, we analyze the observed larger variance in the $T_1$ for Q3. We denote the variance in $T_1$ with $\sigma_{T_1}^2$, and $\sigma_{T_{\mathrm{Purcell}}}^2$ and $\sigma_{T_{\mathrm{others}}}^2$ for the variances in the Purcell decay and non-radiative decay, respectively. We assume that the Purcell and non-radiative decay channels are independent. By computing the uncertainty propagation of \eqref{eq:T1_total} (see Appendix~\ref{app:T_variance} for details), we derive 
\begin{align}
    \sigma_\mathrm{Q3}^2 &\approx \sigma_{T_\mathrm{others}}^2 \\
    \sigma_{\mathrm{Q1/Q2}}^2 &\approx \mleft( \frac{3
    4}{66} \mright)^4 \sigma_{T_{\mathrm{Purcell}}}^2 + \mleft( \frac{34}{70} \mright)^4 \sigma_{T_{\mathrm{others}}}^2\\
        \sigma_{\mathrm{Q4}}^2 &\approx \mleft( \frac{21}{30} \mright)^4 \sigma_{T_{\mathrm{Purcell}}}^2 + \mleft( \frac{21}{70} \mright)^4 \sigma_{T_{\mathrm{others}}}^2,
\end{align}
where the first approximate equality becomes exact if the Purcell decay is completely suppressed. We note from the factor $(34/70)^4 \approx \SI{5.6}{\percent}$ that the contribution from $\sigma_{T_{\mathrm{others}}}^2$ to $\sigma_{Q1/Q4}^2$ is suppressed by about one order of magnitude compared to qubit Q3. We estimate from the experimental data that $\sigma_{\mathrm{Q3}} \approx \SI{30}{\micro\second}$ and average $\sigma_{\mathrm{Q1/Q4}} \approx \SI{4}{\micro\second}$, which gives $\sigma_{T_{\mathrm{Purcell}}} \approx 0$. We note that $\sigma_{\mathrm{Q4}} \approx \SI{7}{\micro\second}$. Therefore, the experimentally observed variations are consistent with small variations in the Purcell decay and that the Purcell decay is strongly suppressed for qubit Q3. 

We note that although both Q2 and Q3 include symmetry-breaking mouse-bite features and exhibit enhanced and fluctuating \( T_1 \) due to engineered interference, the effect is more pronounced for Q3. The partial Purcell suppression in Q2 is attributed to deviations in the experimental assembly. In particular, the separation between Q2’s readout pin and resonator during packaging was smaller than designed, shifting the interference peaks away from the qubit frequency. This misalignment pushed Q2's interference peak to approximately \SI{3}{\giga\hertz} as a complex-frequency transmission zero, resulting in a detuning of \SI{1.84}{\giga\hertz} from Q2's operating frequency of \SI{4.84}{\giga\hertz}. In contrast, Q3 achieved better alignment, with its operating frequency of \SI{5.24}{\giga\hertz} positioned within \SI{340}{\mega\hertz} of its real-frequency interference peak at approximately \SI{4.9}{\giga\hertz}.

\section{Discussion}
\label{sec:Discussion}

The results presented in this article demonstrate that passive multi-mode interference offers a strategy for suppressing Purcell decay in superconducting quantum circuits. By leveraging the inherent complexity of the electromagnetic environment, we introduce a physically grounded design methodology for coherence protection that does not require additional filtering structures or dynamic control circuitry. 

Full-wave eigenmode and admittance simulations confirm that these conditions are physically realisable in practical superconducting layouts. By introducing geometric perturbations that break symmetry---such as a small `mouse-bite' defect---we demonstrate through HFSS simulations and EPR analysis that previously orthogonal modes hybridise, activating interference paths. The increased \(T_1\) observed in the symmetry-broken qubit geometry provides initial experimental validation of this mechanism. 

We estimate the non-radiative decay ceiling at \( T_{1,\text{others}} \approx \SI{70}{\micro\second} \) from reference measurements with external Purcell filters. Our interference-protected qubit (Q3) achieved mean \( T_1 = \SI{66}{\micro\second} \)—a factor of two improvement over the unprotected qubits (\( \sim \SI{30}{\micro\second} \))—while exhibiting significantly larger \( T_1 \) fluctuations (\( \sigma = \SI{30}{\micro\second} \) vs \( \sim \SI{3}{\micro\second} \)). This enhanced variance indicates that Q3 is no longer Purcell-limited, but instead dominated by fluctuating non-radiative decay mechanisms. In contrast, the unprotected qubits show a small variance because their \( T_1 \) is dominated by the Purcell decay rate. This suggests effective Purcell suppression through intrinsic multi-mode interference, with Q3 approaching the non-radiative decay limit. 

Importantly, the strategy for Purcell suppression that we put forward here preserves high gate fidelity and readout compatibility while reducing design complexity. It supports an architectural vision in which coherence emerges not from isolation, but from structured interaction. More broadly, the principles of interference engineering developed here may generalise to other platforms that support multi-mode physics, including photonic, mechanical, and hybrid quantum systems.

Looking forward, this work opens several new directions. Extensions to time-dependent or parametric control of interference channels could enable dynamically reconfigurable filters. Finally, combining this approach with Purcell- and ZZ-coupling-aware layout tools could yield automated design pipelines for large-scale, coherence-optimised quantum processors.

\section*{acknowledgments}
M.B.~acknowledges support from EPSRC QT Fellowship grant EP/W027992/1, and EP/Z53318X/1. P.L.~acknowledges support from EP/N015118/1, and EP/T001062/1. S.P.F.~and A.F.K.~acknowledge support from the Knut and Alice Wallenberg Foundation through the Wallenberg Centre for Quantum Technology (WACQT). A.F.K.~is also supported by the Swedish Foundation for Strategic Research (grant numbers FFL21-0279 and FUS21-0063) and the Horizon Europe programme HORIZON-CL4-2022-QUANTUM-01-SGA via the project 101113946 OpenSuperQPlus100.

\appendix

\section{Network analysis of distributed bandpass filters \label{app:em}}
The outer capacitive pads of the transmon qubit in our device form a ring-shaped structure that supports multiple electromagnetic modes. Understanding these modes is crucial for engineering the multi-mode interference described in the main text. Here, we develop a rigorous analytical framework for describing distributed electromagnetic~\cite{ref51, ref52, ref53} behavior in these circular geometries. 

Consider a conductive ring of radius $R$ as shown in \figref{fig:circuit1}. When the ring's width is small compared to its radius and the wavelength, we can model wave propagation along the ring using a one-dimensional transmission line approximation with an effective propagation constant~\cite{ref30}
\begin{equation}
\beta_{\text{eff}}(\omega) = \frac{\omega}{v_{\text{eff}}},
\label{appendxA:beta_eff}
\end{equation}
where $v_{\text{eff}}$ is the effective phase velocity of the electromagnetic wave in the transmission line. The electromagnetic fields (both current and voltage) vary along the angular coordinate $\theta$ as
\begin{equation}
e^{\mp j\beta_{\text{eff}}R\theta} = e^{\mp j\nu\theta},
\label{appendxA:field_variation}
\end{equation}
where we introduce the dimensionless azimuthal propagation constant
\begin{equation}
\nu(\omega) = \beta_{\text{eff}}(\omega)R = \frac{\omega R}{v_{\text{eff}}}.
\label{appendxA:nu_def}
\end{equation}

This parameter $\nu(\omega)$ is of central importance: it varies continuously with frequency and fully characterizes the angular variation of the fields around the ring. For electromagnetic fields to be single-valued when traversing the ring (i.e., $\theta \rightarrow \theta + 2\pi$), the azimuthal propagation constant must satisfy
\begin{equation}
\nu(\omega_n) = n, \quad n = 0, 1, 2, \ldots
\label{appendxA:quantization_condition}
\end{equation}
This quantization condition leads to resonances at discrete frequencies
\begin{equation}
\omega_n = \frac{n v_{\text{eff}}}{R}, \quad n = 0, 1, 2, \ldots
\label{appendxA:omega_n}
\end{equation}
or equivalently, when the guided wavelength $\lambda_g = 2\pi/\beta_{\text{eff}}$ satisfies
\begin{equation}
2\pi R = n\lambda_g, \quad n = 0, 1, 2, \ldots
\label{appendxA:wavelength_condition}
\end{equation}
These resonance conditions correspond to standing waves around the ring, with electromagnetic fields proportional to $\cos(n\theta)$ and $\sin(n\theta)$. The integer $n$ represents the number of wavelengths that fit around the circumference.

A critical insight for our multi-mode interference framework is that $\nu(\omega)$ is a continuous function of frequency, not merely defined at resonances. At frequencies between resonances, $\nu(\omega)$ takes non-integer values, and the fields exhibit a phase progression around the ring. The general solution for the electromagnetic field distribution at any frequency can be written as
\begin{equation}
\Psi(\theta, \omega) = \sum_n \mleft[ A_n(\omega) e^{j \nu_n(\omega) \theta} + B_n(\omega) e^{-j \nu_n(\omega) \theta} \mright],
\label{appendxA:general_solution}
\end{equation}
where the summation is over all possible modes (values of $n$), $\Psi$ represents either the electric or magnetic field component, and $A(\omega)$ and $B(\omega)$ are complex amplitudes for each mode determined by excitation and boundary conditions. For a ring coupled to external circuits (such as a Josephson junction or control lines), the coefficients $A(\omega)$ and $B(\omega)$ depend on the coupling locations and strengths. With multiple coupling points at different angular positions ${\theta_i}$, the total field becomes a superposition of responses
\begin{equation}
\Psi_{\text{total}}(\theta, \omega) = \sum_i C_i(\omega) \mleft[e^{j\nu_n(\omega)|\theta-\theta_i|} + R_i(\omega)e^{-j\nu_n(\omega)|\theta-\theta_i|}\mright],
\label{appendxA:superposition}
\end{equation}
where $C_i(\omega)$ are amplitude factors and $\mathrm{R}_i(\omega)$ are reflection coefficients at each coupling point. 

Indeed, the outer capacitive pads of our transmon qubits support multiple electromagnetic modes corresponding to different values of $n$. In a symmetric design, modes with different $n$ are orthogonal and do not couple to each other. However, introducing geometric asymmetry (such as the ``mouse-bite'' feature described in Section~\ref{multimodecircuit}) breaks this orthogonality and enables mode mixing. The Josephson junction couples to these modes with strengths that depend on the current distribution at the junction's location. For mode $n$
\begin{equation}
g_n \propto I_n(\theta_J) \propto \cos(n\theta_J),
\label{appendxA:gn}
\end{equation}
where $\theta_J$ is the angular position of the junction. When introducing asymmetry at an angle $\theta_A$, modes with different $n$ values couple with strength
\begin{multline}
J_{nm} \propto \delta(\theta_A) \int_0^{2\pi} \cos(n\theta) \cos(m\theta) 
\delta(\theta - \theta_A)\, d\theta 
\\
\propto \delta(\theta_A) \cos(n\theta_A) \cos(m\theta_A),
\label{appendxA:Jnm}
\end{multline}
where $\delta(\theta_A)$ represents the magnitude of the geometric perturbation. The phase relationship between these mode-mode couplings is crucial for the interference effects described in the main text. 

While \eqref{appendxA:Jnm} provides the coupling magnitudes, the complex phases $\theta_{ij}$ arise from the modal transformation that occurs when geometric asymmetries break the rotational symmetry of the ring geometry. Following the analysis in Ref.~\cite{ref54}, the perturbations hybridize the original orthogonal modes, creating new eigenmodes with modified polarizations and field distributions. The phases $\theta_{ij}$ are determined by the eigenvalue decomposition of the perturbed system and depend on the field overlap integrals of the transformed modes at the coupling location. Hence, by controlling the location $\theta_A$ of the asymmetry relative to the junction position $\theta_J$, we can influence the phases $\phi_i$ and $\theta_{ij}$ in \eqref{eq:multi_mode_H} of the main text, providing a design parameter for engineering interference conditions at the qubit frequency.

\section{Semiclassical analysis: Detailed perturbation expansion of mode amplitudes in $J_{ij}$}
\label{appendixB}

In this appendix, we present the detailed algebraic steps for the semiclassical expansion of the mode amplitudes $a_i$ in powers of the total inter-mode coupling $\sum_{j \neq i} J_{ij}$
compared to detunings, to justify the truncation used in \eqref{eq:qubit_dynamics} of the main text. We explicitly derive the zeroth, first, and second-order terms, and show that the second-order correction is negligible in the regime of weak inter-mode coupling. Starting from \eqref{eq:steady_state_a} in the main text:
\begin{equation}
\label{eq:C1}
a_i = \frac{-ig_i e^{-i\phi_i} \sigma^- - i \sum_{j \ne i} J_{ij} e^{i\theta_{ij}} a_j}{i\Delta_i + \kappa_i/2},
\end{equation}
we define an iterative perturbative solution of the form:
\begin{equation}
\label{eq:C2}
a_i = a_i^{(0)} + a_i^{(1)} + a_i^{(2)} + \cdots,
\end{equation}
where each term $a_i^{(n)}$ is of order $J^n$. 

Setting $J_{ij} = 0$, we obtain the zeroth-order solution:
\begin{equation}
\label{eq:C3}
a_i^{(0)} = \frac{-ig_i e^{-i\phi_i} \sigma^-}{i\Delta_i + \kappa_i/2}.
\end{equation}
Inserting \eqref{eq:C3} into the second term of \eqref{eq:C1}, we find the first-order correction
\begin{equation}
\label{eq:C5}
a_i^{(1)} = \sum_{j \ne i} \frac{g_j J_{ij} e^{i(\theta_{ij} - \phi_j)}}{(i\Delta_i + \kappa_i/2)(i\Delta_j + \kappa_j/2)} \sigma^-.
\end{equation}
To compute the second-order term, we substitute $a_j^{(1)}$ into the sum in \eqref{eq:C1}:
\begin{multline}
\label{eq:C_all}
a_i^{(2)} 
= \frac{-i \sum_{j \ne i} J_{ij} e^{i\theta_{ij}} a_j^{(1)}}{i\Delta_i + \kappa_i/2} 
\\
= \sum_{j \ne i} \sum_{k \ne j} 
\frac{g_k J_{ij} J_{jk} e^{i(\theta_{ij} + \theta_{jk} - \phi_k)}}
{(i\Delta_i + \kappa_i/2)(i\Delta_j + \kappa_j/2)(i\Delta_k + \kappa_k/2)} \sigma^-.
\end{multline}
The second-order term $a_i^{(2)}$ contains products of the form $J_{ij} J_{jk}$, and is therefore of order $\mathcal{O}(J^2)$. In the regime where $\sum_{j \neq i} |J_{ij}| \ll |\Delta_i|, \kappa_i$, these terms are suppressed relative to the leading interference term $\sim J_{ij}$. 

The total correction to the qubit decay rate from second-order paths involves contributions such as:
\begin{multline}
\label{eq:C9}
\Gamma^{(2)} \sim \text{Re}[g_i a_i^{(2)}] 
\\
\propto \sum_{i,j,k} g_i g_k J_{ij} J_{jk} D_{ijk} \cos(\phi_i - \phi_k + \theta_{ij} + \theta_{jk}),
\end{multline}
where $D_{ijk} \sim 1/\Delta_i \Delta_j \Delta_k$ is a denominator of cubic order. Thus, in the perturbative regime $\sum_{j \neq i} |J_{ij}| \ll |\Delta_i|, \kappa_i$, the second-order terms are subleading and do not qualitatively affect the leading-order interference cancellation captured in \eqref{eq:qubit_dynamics} of the main text. In summary, truncating \eqref{eq:steady_state_a} to first order in $J_{ij}$ and substituting only the zeroth-order $a_j$ values yields an expression accurate up to $\mathcal{O}(J)$, which is the dominant contribution to interference effects in the weak coupling regime. The higher-order terms are negligible under the operating conditions of our system and are safely omitted in the main analysis.

\section{Detailed quantum analysis of Purcell decay}
\label{app:density_matrix}
Here, we complement the semiclassical analysis with a quantum-mechanical derivation of the Purcell decay rate. We maintain a tractable analytical derivation by considering the case with no external driving of the system. In this zero-drive regime, we assume that the drive does not excite the resonator, such that the Purcell decay is fully captured within the zero- and single-excitation subspaces. Our analysis of these subspaces follows the one found in Ref.~\cite{ref55}. 

We derive the Purcell decay rate from the Lindblad master equation
\begin{equation}
    \dot{\rho} = -i [H, \rho] + \sum_{i=1}^{m} \mleft( L_i \rho L_i^\dagger - \frac{1}{2} \mleft\{L_i^\dagger L_i, \rho \mright\} \mright) ,
\end{equation}
where $\rho$ is the density matrix of the qubit-multi-mode system, $H$ is the Hamiltonian in \eqref{eq:multi_mode_H} with drive terms set to zero, and $L_i$ is a jump operator for mode $i$. We model the dissipation of each mode as one-excitation relaxation $L_i = \sqrt{\kappa_i} a_i$, where $\kappa_i$ is the relaxation rate for each mode, and neglect the (self-) relaxation of the qubit since we are only interested in the Purcell effect via the modes. We note that the terms $L_i \rho L_i^\dagger$ have no contributions in the single-excitation subspace; they only contribute to the evolution of the zero-excitation state~\cite{ref54}. We can understand this from the fact that $L_i \rho L_i^\dagger$ describes the contribution, e.g., influx of population, from higher to lower excitation subspaces. 

Thus, in the single-excitation subspace, the master equation unravels as $\dot{\rho} = -i [H_\mathrm{eff}, \rho]$, where 
\begin{equation}
    H_\mathrm{eff} = H - \frac{i}{2} \sum \kappa_i a^\dagger_i a_i,
    \label{eq:H_eff}
\end{equation}
is an effective non-Hermitian Hamiltonian governing the non-unitary evolution of the single-excitation subspace. 
We then compute the time evolution from the effective Schrödinger equation $i \partial_t \ket{\psi} = H_\mathrm{eff} \ket{\psi}$. Explicitly, in the single-excitation subspace basis $\{\ket{e,0},\ket{g,1_1},\dots,\ket{g,1_m}\}$, this Hamiltonian has the matrix form, in the lab frame of the qubit,
\begin{equation}
H_{\text{eff}} = 
\begin{pmatrix}
\omega_q & g_1 e^{i\phi_1} & g_2 e^{i\phi_2} & \dots & g_m e^{i\phi_m} \\
g_1e^{-i\phi_1} & \omega_1 - i\frac{\kappa_1}{2} & J_{12} e^{i\theta_{12}} & \dots & J_{1m} e^{i\theta_{1m}}\\
g_2e^{-i\phi_2} & J_{12} e^{-i\theta_{12}} & \omega_2 - i\frac{\kappa_2}{2} & \dots & J_{2m} e^{i\theta_{2m}}\\
\vdots & \vdots & \vdots & \ddots & \vdots\\
g_m e^{-i\phi_m} & J_{1m} e^{-i\theta_{1m}} & J_{2m} e^{-i\theta_{2m}} & \dots & \omega_m - i\frac{\kappa_m}{2}
\end{pmatrix}.
\label{eq:H_eff_lab_fig}
\end{equation}

The Purcell decay is given by the time evolution of the qubit eigenstate $\ket{\psi_e(t)} = c_e(t)\ket{e}$, where $\ket{e}$ is the initial eigenstate with amplitude $c_e(0) = 1$. We solve the effective Schrödinger equation and find $c_e(t) = \exp{(-i \lambda_e t)}$, where $\lambda_e = \omega_e + i\Gamma_e /2$ is the complex-valued eigenenergy of $H_\mathrm{eff}$ that evolves from the bare qubit frequency \( \omega_q \) in the limit of vanishing coupling. It corresponds to the qubit-like eigenstate that dominates the initial condition \( \lvert e \rangle = (1, 0, \ldots, 0)^{\mathrm{T}} \). Here, $\omega_e$ is the eigenfrequency and $\Gamma_e = -2 \operatorname{Im}\lambda_e$ is the corresponding Purcell decay rate. We compute the eigenvalue from third-order perturbation theory:\\
- \textit{Zeroth-order perturbation}: $\omega_q$.\\
- \textit{First-order perturbation}: 0.\\
- \textit{Second-order perturbation}:
\begin{equation}
\lambda_e^{(2)} = \sum_{i=1}^{m}\frac{g_i^2}{\Delta_i + i\frac{\kappa_i}{2}}.
\end{equation}
- \textit{Third-order perturbation} (cross-mode interference terms):
\begin{equation}
\lambda_e^{(3)} = \sum_{i < j} \frac{2 g_i g_j J_{ij} \cos{(\phi_i - \phi_j + \theta_{ij})}}{(\Delta_i + i\frac{\kappa_i}{2})(\Delta_j + i\frac{\kappa_j}{2})}.
\end{equation}

Thus, up to third order, we have:
\begin{equation}
    \lambda_e = \omega_q + \sum_{i=1}^{m}\frac{g_i^2}{\Delta_i + i\frac{\kappa_i}{2}}
    + \sum_{i < j} \frac{2 g_i g_j J_{ij} \cos{(\phi_i - \phi_j + \theta_{ij})}}{(\Delta_i + i\frac{\kappa_i}{2})(\Delta_j + i\frac{\kappa_j}{2})}.
\end{equation}
Extracting the imaginary part, we obtain:
\begin{multline}
\label{qc:purcell_decay_corrected}
\Gamma_e = \sum_{i=1}^{m} 
\frac{\kappa_i g_i^2}{\Delta_i^2 + (\kappa_i/2)^2} 
\\
+ \sum_{i < j} 
\frac{2(\kappa_i\Delta_j + \kappa_j\Delta_i) g_i g_j J_{ij} 
\cos{(\phi_i - \phi_j + \theta_{ij})}}{
(\Delta_i^2 + (\kappa_i/2)^2)(\Delta_j^2 + (\kappa_j/2)^2)}.
\end{multline}
which explicitly matches the semiclassical result in \eqref{eq:purcell_decay}.

\begin{figure}
    \centering
    \includegraphics[width=\linewidth]{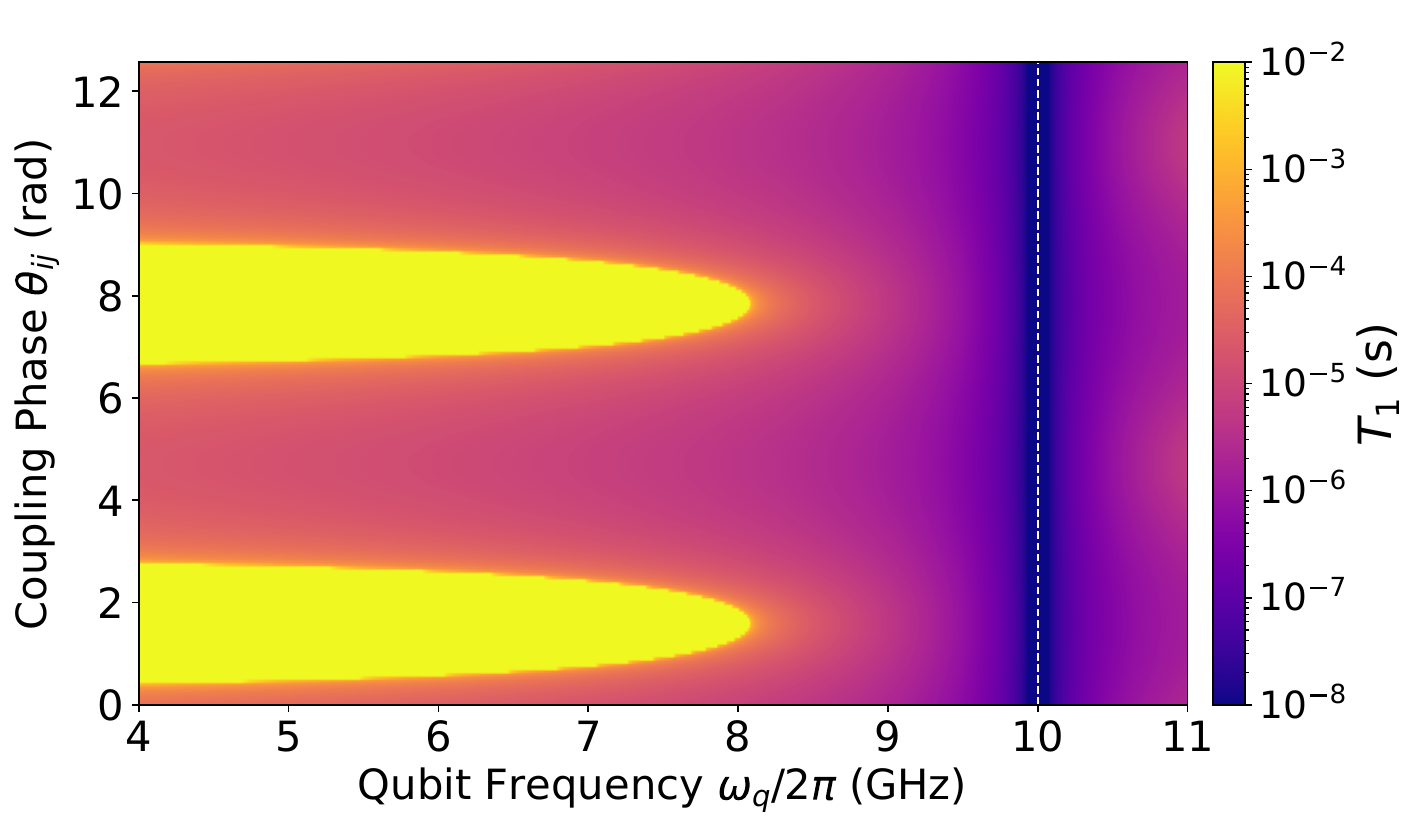}
    \caption{Numerical simulation of qubit relaxation time \( T_1 \) as a function of qubit frequency \( \omega_q \) and coupling phase \( \theta_{ij} \), computed via full diagonalization of the non-Hermitian Hamiltonian \( H_\mathrm{eff} \) in \eqref{eq:H_eff_lab_fig}, capturing both direct decay and interference contributions. The two yellow regions indicate enhanced \( T_1 \) arising from destructive interference between multiple decay pathways. These interference effects are absent in a second-order perturbative approximation, which includes only direct single-mode decay paths, highlighting the critical role of cross-mode interactions. Simulation parameters: \( J_{ij} / 2\pi = \SI{50}{\mega\hertz} \); mode frequencies \( \omega_r / 2\pi = 10 \), 25.05, and \SI{25.08}{\giga\hertz}; and qubit-mode coupling strengths \( g_i / 2\pi = 250,\, 50,\, \text{and} \, \SI{30}{\mega\hertz} \).
    \label{fig:solveeq14}}
\end{figure}

Figure~\ref{fig:solveeq14} demonstrates the numerical validation of our perturbative approach by computing the full diagonalization of the non-Hermitian Hamiltonian \( H_{\text{eff}} \) [\eqref{eq:H_eff}]. The simulation sweeps the qubit frequency \( \omega_q \) and the coupling phase \( \theta_{ij} \) while maintaining representative parameter values for $J_{ij}$, $\omega_i$, and $g_i$, including three modes. The two prominent yellow regions at approximately \qtyrange[range-units=single, range-phrase = --]{4}{8}{\giga\hertz} show enhanced \( T_1 \) values exceeding \SI{e-3}{\second}, arising from destructive interference when the phase conditions align for cancellation. Notably, these interference features are absent in second-order perturbation theory, which includes only direct single-mode decay paths---confirming that the cross-mode coupling terms \( J_{ij} \) are essential for the observed Purcell suppression.

\section{Detailed density-matrix derivation of Purcell decay}
\label{analytical}

In this appendix, we provide yet another derivation of the Purcell decay rate $\Gamma_P$, this time using the density-matrix formalism, explicitly presenting all algebraic steps for clarity. We consider the full system Hamiltonian in \eqref{eq:multi_mode_H} with all external drives set to zero. Dissipation is included through the Lindblad master equation
\begin{equation}
\dot{\rho} = -i[H, \rho] + \sum_i \kappa_i \mathcal{D}[a_i] \rho ,
\end{equation}
where $\kappa_i$ are the photon decay rates of the modes and $\mathcal{D}[a_i] \rho = a_i \rho a_i^\dag - \frac{1}{2} \mleft\{ a_i^\dag a_i, \rho \mright\}$. We restrict our analysis to the single-excitation subspace, spanned by $\ket{e,0}$, where the qubit is excited and all modes are empty, and $\ket{g,1_i}$, where the qubit is in the ground state and mode $i$ contains a single photon. 

Our goal is to compute the decay of the qubit population $\rho_{ee} = \bra{e,0} \rho \ket{e,0}$. Its time evolution follows
\begin{equation}
    \dot{\rho}_{ee} = -\sum_i g_i \, \text{Im}[\rho_{e,1_i}] - \sum_i \kappa_i \rho_{1_i,1_i},
    \label{eq:rho_ee_dot}
\end{equation}
where $\rho_{e,1_i} = \bra{e,0} \rho \ket{g,1_i}$ describes the coherence between the excited qubit and single-photon mode states, and $\rho_{1_i,1_j} = \bra{g,1_i} \rho \ket{g,1_j}$ accounts for coherences between different modes. Under the assumption that the resonator modes decay much faster than the qubit evolves, any photons that leak from the qubit into the resonator modes are quickly dissipated to the environment before they can accumulate. This Born--Markov approximation ensures that $\rho_{ee} \gg \rho_{1_i,1_j}$ during the qubit evolution, justifying neglecting the mode populations $\rho_{1_i,1_i}$ in \eqref{eq:rho_ee_dot}. This approximation simplifies the evolution of $\rho_{e,1_i}$ to
\begin{equation}
    \dot{\rho}_{e,1_i} \approx -i \Delta_i \rho_{e,1_i} - i g_i \rho_{ee} - i \sum_{j \neq i} J_{ij} \rho_{e,1_j} - \frac{\kappa_i}{2} \rho_{e,1_i},
    \label{eq:rho_e1i_weak}
\end{equation}
where $\Delta_i = \omega_i - \omega_q$ is the detuning between the qubit and mode $i$ in the rotating frame of the qubit. 

We introduce the shorthand notation $D_i \equiv i \Delta_i + \kappa_i / 2$ and note that under the Born--Markov approximation, the coherences \(\rho_{e,1_i}\) evolve much faster than the qubit population \(\rho_{ee}\). Thus, we can set \(\dot{\rho}_{e,1_i} = 0\) in \eqref{eq:rho_e1i_weak}, yielding the algebraic relation
\begin{equation}
    D_i \, \rho_{e,1_i} = -i g_i \rho_{ee} - i \sum_{j \neq i} J_{ij} \rho_{e,1_j}.
    \label{eq:base_eq}
\end{equation}
This equation reveals two contributions: the first term describes the direct qubit-mode coupling, while the second term captures cross-mode effects arising from the coherent coupling between modes. Solving for $\rho_{e,1_i}$, we obtain
\begin{equation}
    \rho_{e,1_i} = \frac{-i g_i \rho_{ee}}{D_i} + \sum_{j \neq i} \frac{-i J_{ij}}{D_i} \rho_{e,1_j}.
    \label{eq:rho_e1i_formal}
\end{equation}

To solve this recursive expression, we substitute $\rho_{e,1_j}$ using the same form:
\begin{align}
    \rho_{e,1_i} &= \frac{-i g_i \rho_{ee}}{D_i} + \sum_{j \neq i} \frac{-i J_{ij}}{D_i} \mleft( \frac{-i g_j \rho_{ee}}{D_j} + \sum_{k \neq j} \frac{-i J_{jk}}{D_j} \rho_{e,1_k} \mright) \nonumber \\
    &= \frac{-i g_i \rho_{ee}}{D_i} + \sum_{j \neq i} \frac{J_{ij} g_j \rho_{ee}}{D_i D_j} + \sum_{j \neq i} \sum_{k \neq j} \frac{J_{ij} J_{jk}}{D_i D_j} \rho_{e,1_k}.
    \label{eq:recursion1}
\end{align}
Applying recursive substitution once more, we express $\rho_{e,1_k}$ in the same way and substitute:
\begin{widetext}
\begin{align}
    \rho_{e,1_i} &= \frac{-i g_i \rho_{ee}}{D_i} + \sum_{j \neq i} \frac{J_{ij} g_j \rho_{ee}}{D_i D_j} + \sum_{j \neq i} \sum_{k \neq j} \frac{J_{ij} J_{jk}}{D_i D_j} \mleft( \frac{-i g_k \rho_{ee}}{D_k} + \sum_{l \neq k} \frac{-i J_{kl}}{D_k} \rho_{e,1_l} \mright) \nonumber \\
    &= \frac{-i g_i \rho_{ee}}{D_i} + \sum_{j \neq i} \frac{J_{ij} g_j \rho_{ee}}{D_i D_j} - i \sum_{j \neq i} \sum_{k \neq j} \frac{J_{ij} J_{jk} g_k \rho_{ee}}{D_i D_j D_k} + \sum_{j \neq i} \sum_{k \neq j} \sum_{l \neq k} \frac{J_{ij} J_{jk} J_{kl}}{D_i D_j D_k} \rho_{e,1_l}.
    \label{eq:recursion2}
\end{align}
The final term represents contributions at order $\mathcal{O}[(J/D)^3]$, which we neglect assuming weak mode-mode couplings $\sum_{j \neq i} |J_{ij}|/|D_i| \ll 1$, where \( D_i = i\Delta_i + \kappa_i/2 \). Truncating at second order in the small parameter $\sum J/D$, we obtain
\begin{equation}
    \rho_{e,1_i} \approx \frac{-i g_i \rho_{ee}}{D_i} + \sum_{j \neq i} \frac{J_{ij} g_j \rho_{ee}}{D_i D_j} - i \sum_{j \neq i} \sum_{k \neq j} \frac{J_{ij} J_{jk} g_k \rho_{ee}}{D_i D_j D_k}.
    \label{eq:rho_e1i_truncated}
\end{equation}

We now substitute \eqref{eq:rho_e1i_truncated} into the expression for $\dot{\rho}_{ee}$ from \eqref{eq:rho_ee_dot}. Retaining terms up to $\mathcal{O}[(J/D)^2]$, the evaluation of the imaginary part yields
\begin{align}
\dot{\rho}_{ee} &= -\sum_i g_i \, \text{Im} \mleft[ \frac{-i g_i \rho_{ee}}{D_i} 
+ \sum_{j \neq i} \frac{J_{ij} g_j \rho_{ee}}{D_i D_j} 
- i \sum_{j \neq i} \sum_{k \neq j} \frac{J_{ij} J_{jk} g_k \rho_{ee}}{D_i D_j D_k} \mright] \nonumber \\
&= \rho_{ee} \sum_i \frac{g_i^2 \kappa_i}{\Delta_i^2 + (\kappa_i/2)^2} 
+ \rho_{ee} \sum_{i \neq j} \frac{(\kappa_i \Delta_j + \kappa_j \Delta_i) g_i g_j J_{ij}}{(\Delta_i^2 + (\kappa_i/2)^2)(\Delta_j^2 + (\kappa_j/2)^2)} \nonumber \\
&\quad- \rho_{ee} \sum_{i \neq j} \sum_{k \neq j} \frac{g_i J_{ij} J_{jk} g_k (\kappa_i \Delta_k + \kappa_k \Delta_i)}{(\Delta_i^2 + (\kappa_i/2)^2)(\Delta_j^2 + (\kappa_j/2)^2)(\Delta_k^2 + (\kappa_k/2)^2)}.
\label{eq:gamma_purcell_full}
\end{align}
\end{widetext}
The first term corresponds to direct single-mode Purcell decay, while the second term captures interference between decay paths through different modes. The third term arises from three-mode interference and scales as $\mathcal{O}[(J/D)^2]$, remaining small for weak mode-mode couplings. Finally, we extract the effective decay rate by identifying $\dot{\rho}_{ee} = -\Gamma_P \rho_{ee}$, leading to
\begin{equation}
    \Gamma_P \approx \sum_i \frac{g_i^2 \kappa_i}{\Delta_i^2 + (\kappa_i/2)^2} - \sum_{i \neq j} \frac{(\kappa_i \Delta_j + \kappa_j \Delta_i) g_i g_j J_{ij}}{(\Delta_i^2 + (\kappa_i/2)^2)(\Delta_j^2 + (\kappa_j/2)^2)}.
    \label{eq:gamma_purcell_final}
\end{equation}
This result matches the expressions obtained from both semiclassical analysis [\eqref{eq:purcell_decay_corrected}] and non-Hermitian perturbation theory [\eqref{qc:purcell_decay_corrected}], providing a consistent picture of Purcell decay and its suppression via multi-mode quantum interference.

\section{Derivation of the cross-Kerr coefficients}
\label{crossker}
In this appendix, we derive the cross-Kerr interaction coefficients $\chi_{kl}$ that contribute to the AC Stark shift of the qubit under coherent drive in a multi-mode system, using the simplified two-level qubit approximation. Note that the dispersive shifts $\chi_k$ used in the main text (Section~\ref{sec:AnalysisDrivenPurcell}) include anharmonic corrections from the full multi-level transmon, which require virtual transitions through $\ket{2}$, $\ket{3}$, etc. The complete multi-level derivation is beyond the scope of this work and can be found in standard cQED references~\cite{ref38}. The starting point is the Hamiltonian in the lab frame and diagonalized basis
\begin{equation}
H = \omega_q \sigma^+ \sigma^- + \sum_k \tilde{\omega}_k \tilde{a}^\dagger_k \tilde{a}_k + \sum_k \mleft( \tilde{g}_k \tilde{a}_k^\dagger \sigma^- + \tilde{g}_k^* \tilde{a}_k \sigma^+ \mright),
\label{eq:app_hamiltonian}
\end{equation}
where $\tilde{a}_k$ are the annihilation operators of the diagonalized normal modes with frequencies $\tilde{\omega}_k$, and $\tilde{g}_k \in \mathbb{C}$ are the effective couplings of these modes to the qubit. We define the detunings
\begin{equation}
\tilde{\Delta}_k = \omega_q - \tilde{\omega}_k.
\label{eq:app_detuning}
\end{equation}

In the dispersive limit $|\tilde{g}_k / \tilde{\Delta}_k| \ll 1$, we apply a Schrieffer--Wolff transformation to eliminate the qubit-mode interaction to second order. We define the generator
\begin{equation}
S = \sum_k \mleft( \frac{\tilde{g}_k}{\tilde{\Delta}_k} \tilde{a}_k^\dagger \sigma^- - \frac{\tilde{g}_k^*}{\tilde{\Delta}_k} \tilde{a}_k \sigma^+ \mright),
\label{eq:app_generator}
\end{equation}
which yields an effective Hamiltonian
\begin{equation}
H_{\text{eff}} = e^{-S} H e^S \approx H + [H, S] + \frac{1}{2}[[H, S], S].
\label{eq:app_effham}
\end{equation}

We now compute the nested commutators. The relevant second-order term that contributes to the qubit-state-dependent cross-Kerr interaction is
\begin{equation}
H_{\chi} = \frac{1}{2} \mleft[ \mleft[H_{\text{int}}, S \mright], S \mright],
\label{eq:app_chiterm}
\end{equation}
where $H_{\text{int}} = \sum_k \mleft( \tilde{g}_k \tilde{a}_k^\dagger \sigma^- + \text{H.c.} \mright)$. Evaluating this gives a cross-Kerr term of the form
\begin{multline}
\label{eq:Hchi_full}
H_{\chi} = 
\sum_k \chi_k \tilde{a}^\dagger_k \tilde{a}_k \, \sigma^+ \sigma^-
+ \sum_{k \ne l} \chi_{kl} \tilde{a}^\dagger_k \tilde{a}_k \tilde{a}^\dagger_l \tilde{a}_l \, \sigma^+ \sigma^-
\\
+ \sum_k \delta_k \sigma^+ \sigma^-
+ \sum_k \delta \tilde{\omega}_k \tilde{a}^\dagger_k \tilde{a}_k
+ \sum_{k \ne l} \eta_{kl} \tilde{a}^\dagger_k \tilde{a}_l
\end{multline}
with
\begin{equation}
\chi_{kl} = 
\begin{cases}
\displaystyle -\frac{|\tilde{g}_k|^2}{\tilde{\Delta}_k} & \text{if } k = l, \\[10pt]
\displaystyle -\frac{\tilde{g}_k^* \tilde{g}_l}{2} \mleft( \frac{1}{\tilde{\Delta}_k} + \frac{1}{\tilde{\Delta}_l} \mright) & \text{if } k \neq l,
\end{cases}
\label{eq:app_chikl}
\end{equation}
where $\delta_k$ is the Lamb shift of the qubit, $\delta \tilde{\omega}_k$ is the renormalized mode frequency (back-action from virtual qubit excitation) in the frame rotating with the bare qubit frequency, and $\eta_{kl}$ is the qubit-mediated coherent mode-mode exchange term. The case $k = l$ corresponds to the self-Kerr (usual AC Stark shift), while $k \neq l$ gives the cross-Kerr interaction between different modes due to the shared qubit nonlinearity. 

The cross terms can be significant when multiple modes are simultaneously populated. The total qubit frequency shift in the presence of coherent populations $\bar{n}_k = \langle \tilde{a}_k^\dagger \tilde{a}_k \rangle$ is then
\begin{equation}
\omega_{q,\text{eff}} = \omega_q + \sum_k 2 \chi_{kk} \bar{n}_k + \sum_{k \ne l} 2 \chi_{kl} \bar{n}_k \bar{n}_l + \mathcal{O}(\tilde{g}^4/\tilde{\Delta}^3).
\label{eq:app_wq_eff}
\end{equation}
Note that these expressions for the self-Kerr and cross-Kerr coefficients $\chi_{kl}$ feed directly into the AC Stark shift of the qubit frequency, as discussed in Section~\ref{sec:AnalysisDrivenPurcell}. In particular, the effective detunings $\tilde{\Delta}_{k,\text{eff}}(\bar{n})$ incorporate these Kerr terms through the drive-induced shift of the qubit transition frequency
\begin{equation}
\tilde{\Delta}_{k,\text{eff}}(\bar{n}) = \tilde{\Delta}_k + 2\chi_{\text{eff}}\bar{n},
\label{appE: effective_detuning}
\end{equation}
where $\chi_{\text{eff}}$ is a weighted sum of the Kerr coefficients $\chi_k = \chi_{kk}$ and $\chi_{kl}$ across all populated normal modes. This connection explains how the photon occupation modifies both direct and interference contributions to the Purcell decay rate, and why the interference term exhibits enhanced suppression as a function of $\bar{n}$ in \figref{n_purcell_decay_rate}.

\section{Notes on $T_1$ variance from two independent decay channels}
\label{app:T_variance}
Suppose that we have two \emph{independent} and fluctuating decay channels with decay rates $\Gamma_a$ and $\Gamma_b$, which we assume are stochastic variables. The independent property implies for $\Gamma_\mathrm{tot} = \Gamma_a + \Gamma_b$:
\begin{equation}
    \mathrm{Var}(\Gamma_\mathrm{tot}) = \mathrm{Var}(\Gamma_a) + \mathrm{Var}(\Gamma_b).
    \label{eq:variance}
\end{equation}

We now assume that we can assign a (fluctuating) time $T$ to each decay rate such that $\Gamma = 1/T$. We also assume that $T$ fluctuates around a mean $\bar{T}$ with fluctuations $\delta T = T - \bar{T}$ smaller than the mean. If $|\delta T / \bar{T}| \ll 1$, we can expand the nonlinear relation $\Gamma = 1/T$ to first order in this small parameter:
\begin{equation}
    \Gamma \approx \frac{1}{\bar{T}} \mleft(1 - \frac{\delta T}{\bar{T}} \mright).
\end{equation}
Inserting this approximation into \eqref{eq:variance} yields
\begin{equation}
    \frac{1}{\bar{T}_\mathrm{tot}^4} \mathrm{Var}(T_\mathrm{tot}) = \frac{1}{\bar{T}_a^4} \mathrm{Var}(T_a) + \frac{1}{\bar{T}_b^4} \mathrm{Var}(T_b),
\end{equation}
where we used $\mathrm{Var}(\delta T) = \mathrm{Var}(T)$ and the variance property $\mathrm{Var}(kT+C) = k^2 \mathrm{Var}(T)$ for constants $k$ and $C$.

\bibliographystyle{nature}

\begin{thebibliography}{99}

\bibitem{ref1}
A.~Megrant and Y.~Chen, ``Scaling up superconducting quantum computers,'' \href{https://doi.org/10.1038/s41928-025-01381-7}{\emph{Nature Electronics} \textbf{8} (2025)}.

\bibitem{ref2}
R.~Acharya, I.~Aleiner, R.~Allen \emph{et al.}, ``Suppressing quantum errors by scaling a surface code logical qubit,'' 
\href{https://doi.org/10.1038/s41586-022-05434-1}{\emph{Nature} \textbf{614}, 676--681 (2023)}.

\bibitem{ref3}
M.~Kjaergaard, M.~E.~Schwartz, J.~Braumüller \emph{et al.}, ``Superconducting qubits: current state of play,'' 
\href{https://doi.org/10.1146/annurev-conmatphys-031119-050605}{\emph{Annu. Rev. Condens. Matter Phys.} \textbf{11}, 369--395 (2020)}.

\bibitem{ref4}
A.~Blais, A.~L.~Grimsmo, S.~M.~Girvin \emph{et al.}, ``Circuit quantum electrodynamics,'' 
\href{https://doi.org/10.1103/RevModPhys.93.025005}{\emph{Rev. Mod. Phys.} \textbf{93}, 025005 (2021)}.

\bibitem{ref5}
X.~Gu, A.~F.~Kockum, A.~Miranowicz \emph{et al.}, ``Microwave photonics with superconducting quantum circuits,'' 
\href{https://doi.org/10.1016/j.physrep.2017.10.002}{\emph{Phys. Rep.} \textbf{718--719}, 1--102 (2017)}.

\bibitem{ref6}
A.~Wallraff, D.~I.~Schuster, A.~Blais \emph{et al.}, ``Strong coupling of a single photon to a superconducting qubit using circuit quantum electrodynamics,'' 
\href{https://doi.org/10.1038/nature02851}{\emph{Nature} \textbf{431}, 162--167 (2004)}.

\bibitem{ref7}
A.~Blais, R.~S.~Huang, A.~Wallraff \emph{et al.}, ``Cavity quantum electrodynamics for superconducting electrical circuits: An architecture for quantum computation,'' 
\href{https://doi.org/10.1103/PhysRevA.69.062320}{\emph{Phys. Rev. A} \textbf{69}, 062320 (2004)}.

\bibitem{ref8}
P.~Krantz, M.~Kjaergaard, F.~Yan \emph{et al.}, ``A quantum engineer's guide to superconducting qubits,'' 
\href{https://doi.org/10.1063/1.5089550}{\emph{Appl. Phys. Rev.} \textbf{6}, 021318 (2019)}.


\bibitem{ref9}
S.~Cao, Z.~Shao, J.-Q.~Zheng \emph{et al.}, ``Superconducting qubit readout enhanced by path signature,'' 
\href{https://arxiv.org/abs/2402.09532}{arXiv:2402.09532 [quant-ph]} (2025).


\bibitem{ref10}
E. M. Purcell,
``Proceedings of the American Physical Society,'' 
\href{https://doi.org/10.1103/PhysRev.69.674.2}{\emph{Phys. Rev.} \textbf{69}, 674 (1946)}.


\bibitem{ref11}
R.~Bianchetti, S.~Filipp, M.~Baur \emph{et al.}, ``Dynamics of dispersive single-qubit readout in circuit quantum electrodynamics,'' 
\href{https://doi.org/10.1103/PhysRevA.80.043840}{\emph{Phys. Rev. A} \textbf{80}, 043840 (2009)}. 


\bibitem{ref12}
B.~M.~Smitham, J.~G.~C.~Martinez, C.~S.~Chiu \emph{et al.}, ``Sub-resonant wideband superconducting Purcell filters,'' 
\href{https://arxiv.org/abs/2503.10750}{arXiv:2503.10750 [quant-ph]} (2025).


\bibitem{ref13}
Y.~Zhou, X.~Cai, Y.~Zheng \emph{et al.}, ``High-suppression-ratio and wide bandwidth four-stage Purcell filter for multiplexed superconducting qubit readout,'' 
\href{https://doi.org/10.1063/5.0173539}{\emph{J. Appl. Phys.} \textbf{135}, 024402 (2024)}.


\bibitem{ref14}
H.~Yan, X.~Wu, A.~Lingenfelter \emph{et al.}, ``Broadband bandpass Purcell filter for circuit quantum electrodynamics,'' 
\href{https://doi.org/10.1063/5.0161893}{\emph{Appl. Phys. Lett.} \textbf{123}, 134001 (2023)}.


\bibitem{ref15}
P.~Patel, M.~Xia, C.~Zhou \emph{et al.}, ``The waves-in-space Purcell effect for superconducting qubits,'' 
\href{https://arxiv.org/abs/2503.11644}{arXiv:2503.11644 [quant-ph]} (2025).


\bibitem{ref16}
A.~Yen, Y.~Ye, K.~Peng \emph{et al.}, ``Interferometric Purcell suppression of spontaneous emission in a superconducting qubit,'' 
\href{https://doi.org/10.1103/PhysRevApplied.23.024068}{\emph{Phys. Rev. Appl.} \textbf{23}, 024068 (2025)}.


\bibitem{ref17}
Y.~Sunada, S.~Kono, J.~Ilves \emph{et al.}, ``Fast readout and reset of a superconducting qubit coupled to a resonator with an intrinsic Purcell filter,'' 
\href{https://doi.org/10.1103/PhysRevApplied.17.044016}{\emph{Phys. Rev. Appl.} \textbf{17}, 044016 (2022)}.


\bibitem{ref18}
J.~M.~Gambetta, A.~A.~Houck, A.~Blais, ``Superconducting qubit with Purcell protection and tunable coupling,'' 
\href{https://doi.org/10.1103/PhysRevLett.106.030502}{\emph{Phys. Rev. Lett.} \textbf{106}, 030502 (2011)}.

\bibitem{ref19}
M.~D.~Reed, B.~R.~Johnson, A.~A.~Houck \emph{et al.}, ``Fast reset and suppressing spontaneous emission of a superconducting qubit,'' 
\href{https://doi.org/10.1063/1.3435463}{\emph{Appl. Phys. Lett.} \textbf{96}, 203110 (2010)}.

\bibitem{ref20}
P.~A.~Spring, L.~Milanovic, Y.~Sunada \emph{et al.}, ``Fast multiplexed superconducting qubit readout with intrinsic Purcell filtering,'' 
\href{https://arxiv.org/abs/2409.04967}{arXiv:2409.04967 [quant-ph]} (2024).

\bibitem{ref21}
S.~Kono, K.~Koshino, D.~Lachance-Quirion \emph{et al.}, ``Breaking the trade-off between fast control and long lifetime of a superconducting qubit,'' 
\href{https://doi.org/10.1038/s41467-020-17511-y}{\emph{Nat. Commun.} \textbf{11}, 3683 (2020)}.

\bibitem{ref22}
K.~Koshino, S.~Kono, Y.~Nakamura, ``Protection of a qubit via subradiance: A Josephson quantum filter,'' 
\href{https://doi.org/10.1103/PhysRevApplied.13.014051}{\emph{Phys. Rev. Appl.} \textbf{13}, 014051 (2020)}.

\bibitem{ref23}
J.~Hu, D.~Li, Y.~Qie \emph{et al.}, ``Engineering the environment of a superconducting qubit with an artificial giant atom,'' 
\href{https://arxiv.org/abs/2410.15377}{arXiv:2410.15377 [quant-ph]} (2024).

\bibitem{ref24}
M.~Bakr, S.~D.~Fasciati, S.~Cao \emph{et al.}, ``Multiplexed readout of superconducting qubits using a three-dimensional reentrant-cavity filter,'' 
\href{https://doi.org/10.1103/PhysRevApplied.23.054089}{\emph{Phys. Rev. Appl.} \textbf{23}, 054089 (2025)}.

\bibitem{ref25}
J.~Heinsoo, C.~K.~Andersen, A.~Remm \emph{et al.}, ``Rapid high-fidelity multiplexed readout of superconducting qubits,'' 
\href{https://doi.org/10.1103/PhysRevApplied.10.034040}{\emph{Phys. Rev. Appl.} \textbf{10}, 034040 (2018)}.

\bibitem{ref26}
S.~Kundu, N.~Gheeraert, S.~Hazra \emph{et al.}, ``Multiplexed readout of four qubits in 3D circuit QED architecture using a broadband Josephson parametric amplifier,'' 
\href{https://doi.org/10.1063/1.5089729}{\emph{Appl. Phys. Lett.} \textbf{114}, 172601 (2019)}.

\bibitem{ref27}
M.~S.~Bakr, ``Triple-mode microwave filters with arbitrary prescribed transmission zeros,'' 
\href{https://doi.org/10.1109/ACCESS.2021.3052059}{\emph{IEEE Access} \textbf{9}, 22045--22052 (2021)}.

\bibitem{ref28}
M.~S.~Bakr, I.~C.~Hunter, W.~Bösch, ``Miniature triple-mode dielectric resonator filters,'' 
\href{https://doi.org/10.1109/TMTT.2018.2873309}{\emph{IEEE Trans. Microw. Theory Techn.} \textbf{66}, 5625--5631 (2018)}.

\bibitem{ref29}
M.~S.~Bakr, S.~W.~O.~Luhaib, I.~C.~Hunter \emph{et al.}, ``Dual-mode dual-band conductor-loaded dielectric resonator filters,'' 
\href{https://doi.org/10.23919/EuMC.2017.8230992}{in \emph{Proc. 47th Eur. Microw. Conf. (EuMC)}, 908--910 (2017)}.

\bibitem{ref30}
M.~Bakr and S.~Amari, ``Singular Azimuthally Propagating Electromagnetic Fields,'' 
\href{https://doi.org/10.48550/arXiv.2305.08869}{{arXiv:2305.08869} [physics.class-ph]} (2023).


\bibitem{ref31}
S.~Ahmadi, M.~Akbari, S.~Saeidian \emph{et al.}, \emph{Quantum interference in atomic systems}, 
\href{https://arxiv.org/abs/2502.01398}{arXiv:2502.01398 [quant-ph]} (2025).

\bibitem{ref32}
A.~F.~Kockum, ``Quantum optics with giant atoms—the first five years,'' 
in \emph{Int. Symp. on Mathematics, Quantum Theory, and Cryptography}, 
T.~Takagi \emph{et al.}, Eds., Springer, Singapore, pp.~125--146 (2021). 
\href{https://doi.org/10.1007/978-981-15-5191-8_12}{doi:10.1007/978-981-15-5191-8\_12}.

\bibitem{ref33}
M.~Alghadeer, S.~Cao, S.~D.~Fasciati \emph{et al.}, ``Low crosstalk in a scalable superconducting quantum lattice,'' 
\href{https://arxiv.org/abs/2505.22276}{arXiv:2505.22276 [quant-ph]} (2025).


\bibitem{ref34}
S.~Pettersson Fors, J.~Fernández-Pendás, A.~F.~Kockum, ``Comprehensive explanation of ZZ coupling in superconducting qubits,'' 
\href{https://arxiv.org/abs/2408.15402}{arXiv:2408.15402 [quant-ph]} (2024).

\bibitem{ref35}
J.~M.~Chow, A.~D.~Córcoles, J.~M.~Gambetta \emph{et al.}, ``Simple all-microwave entangling gate for fixed-frequency superconducting qubits,'' 
\href{https://doi.org/10.1103/PhysRevLett.107.080502}{\emph{Phys. Rev. Lett.} \textbf{107}, 080502 (2011)}.


\bibitem{ref36}
Y.~Sung, L.~Ding, J.~Braumüller \emph{et al.}, ``Realization of high-fidelity CZ and $ZZ$-free iSWAP gates with a tunable coupler,'' 
\href{https://doi.org/10.1103/PhysRevX.11.021058}{\emph{Phys. Rev. X} \textbf{11}, 021058 (2021)}.


\bibitem{ref37}
J.~Chu and F.~Yan, ``Coupler-assisted controlled-phase gate with enhanced adiabaticity,'' 
\href{https://doi.org/10.1103/PhysRevApplied.16.054020}{\emph{Phys. Rev. Appl.} \textbf{16}, 054020 (2021)}.

\bibitem{ref38}
J.~Koch, T.~M.~Yu, J.~Gambetta \emph{et al.}, 
``Charge-insensitive qubit design derived from the Cooper pair box,'' 
\href{https://doi.org/10.1103/PhysRevA.76.042319}{\emph{Phys. Rev. A} \textbf{76}, 042319 (2007)}.


\bibitem{ref39}
A.~A.~Houck, J.~A.~Schreier, B.~R.~Johnson \emph{et al.}, ``Controlling the spontaneous emission of a superconducting transmon qubit,'' 
\href{https://doi.org/10.1103/PhysRevLett.101.080502}{\emph{Phys. Rev. Lett.} \textbf{101}, 080502 (2008)}.


\bibitem{ref40}
J.~M.~Fink, M.~Göppl, M.~Baur \emph{et al.}, ``Climbing the Jaynes–Cummings ladder and observing its nonlinearity in a cavity QED system,'' 
\href{https://doi.org/10.1038/nature07112}{\emph{Nature} \textbf{454}, 315--318 (2008)}.


\bibitem{ref41}
A.~Blais, R.~S.~Huang, A.~Wallraff \emph{et al.}, ``Cavity quantum electrodynamics for superconducting electrical circuits: An architecture for quantum computation,'' 
\href{https://doi.org/10.1103/PhysRevA.69.062320}{\emph{Phys. Rev. A} \textbf{69}, 062320 (2004)}.


\bibitem{ref42}
P.~A.~Spring, T.~Tsunoda, B.~Vlastakis \emph{et al.}, ``Modeling enclosures for large-scale superconducting quantum circuits,'' 
\href{https://doi.org/10.1103/PhysRevApplied.14.024061}{\emph{Phys. Rev. Appl.} \textbf{14}, 024061 (2020)}.

\bibitem{ref43}
M.~Bakr and S.~Amari, ``Theory of azimuthally propagating electromagnetic waves in cylindrical cavities,'' 
\href{https://arxiv.org/abs/2505.04756}{arXiv:2505.04756 [physics.optics]} (2025).

\bibitem{ref44}
Z.~K.~Minev, Z.~Leghtas, S.~O.~Mundhada \emph{et al.}, ``Energy-participation quantization of Josephson circuits,'' 
\href{https://doi.org/10.1038/s41534-021-00461-8}{\emph{npj Quantum Inf.} \textbf{7}, 131 (2021)}.


\bibitem{ref45}
M.~Alghadeer, S.~D.~Fasciati, S.~Cao \emph{et al.}, ``Characterization of nanostructural imperfections in superconducting quantum circuits,'' 
\href{https://arxiv.org/abs/2501.15059}{arXiv:2501.15059 [quant-ph]} (2025).

\bibitem{Soro2023}
A.~Soro, C.~S.~Mu\~noz, A.~F.~Kockum, ``Interaction between giant atoms in a one-dimensional structured environment,'' 
\href{https://doi.org/10.1103/PhysRevA.107.013710}{\emph{Phys. Rev. A} \textbf{107}, 013710 (2023)}.

\bibitem{Ingelsten2024}
E.~R.~Ingelsten, A.~F.~Kockum, A.~Soro, ``Avoiding decoherence with giant atoms in a two-dimensional structured environment,'' 
\href{https://doi.org/10.1103/PhysRevResearch.6.043222}{\emph{Phys. Rev. Res.} \textbf{6}, 043222 (2024)}.

\bibitem{ref46}
S.~D.~Fasciati, B.~Shteynas, G.~Campanaro \emph{et al.}, ``Complementing the transmon by integrating a geometric shunt inductor,'' 
\href{https://arxiv.org/abs/2410.10416}{arXiv:2410.10416 [quant-ph]} (2024).


\bibitem{ref47}
E.~A.~Sete, J.~M.~Gambetta, A.~N.~Korotkov, ``Purcell effect with microwave drive: Suppression of qubit relaxation rate,'' 
\href{https://doi.org/10.1103/PhysRevB.89.104516}{\emph{Phys. Rev. B} \textbf{89}, 104516 (2014)}.


\bibitem{ref48}
P.~A.~Spring, S.~Cao, T.~Tsunoda \emph{et al.}, ``High coherence and low cross-talk in a tileable 3D integrated superconducting circuit architecture,'' 
\href{https://doi.org/10.1126/sciadv.abl6698}{\emph{Sci. Adv.} \textbf{8}, eabl6698 (2022)}.


\bibitem{ref49}
M.~Bakr, ``Dynamic Josephson junction metasurfaces for multiplexed control of superconducting qubits,'' 
\href{https://arxiv.org/abs/2411.01345}{arXiv:2411.01345} [quant-ph] (2024).

\bibitem{cao2024}
S.~Cao, M.~Bakr, G.~Campanaro, S.~D.~Fasciati, J.~Wills, D.~Lall, B.~Shteynas, V.~Chidambaram, I.~Rungger, and P.~Leek, ``Emulating two qubits with a four-level transmon qudit for variational quantum algorithms,'' 
\href{https://doi.org/10.1088/2058-9565/ad37d4}{\emph{Quantum Sci. Technol.} \textbf{9}, 035003 (2024)}.


\bibitem{ref50}
F.~Motzoi, J.~M.~Gambetta, P.~Rebentrost, and F.~K.~Wilhelm, ``Simple Pulses for Elimination of Leakage in Weakly Nonlinear Qubits,'' 
\href{https://doi.org/10.1103/PhysRevLett.103.110501}{\emph{Phys. Rev. Lett.} \textbf{103}, 110501 (2009)}.


\bibitem{ref51}
E.~Musonda and M.~S.~Bakr, ``The Singlet: Direct Synthesis of Pseudo-Elliptic Inline Filters With Frequency Variant Couplings,'' 
\href{https://doi.org/10.1109/TMTT.2023.3269516}{\emph{IEEE Trans. Microw. Theory Techn.} \textbf{71}, 4969--4981 (2023)}.


\bibitem{ref52}
S.~Amari, M.~Bakr, and U.~Rosenberg, ``Properties of building blocks comprising strongly interacting posts and their consideration in advanced coaxial filter designs,'' 
\href{https://arxiv.org/abs/2505.15729}{arXiv:2505.15729 [physics.class-ph]} (2025).


\bibitem{ref53}
F.~Solgun, D.~W.~Abraham, and D.~P.~DiVincenzo, ``Blackbox quantization of superconducting circuits using exact impedance synthesis,'' 
\href{https://doi.org/10.1103/PhysRevB.90.134504}{\emph{Phys. Rev. B} \textbf{90}, 134504 (2014)}.


\bibitem{ref54}
M.~Bekheit and S.~Amari, ``A direct design technique for dual-mode inline microwave bandpass filters,'' 
\href{https://doi.org/10.1109/TMTT.2009.2027165}{\emph{IEEE Trans. Microw. Theory Techn.} \textbf{57}, 2193--2202 (2009)}.

\bibitem{ref55}
E.~A.~Sete, J.~M.~Martinis, and A.~N.~Korotkov, ``Quantum theory of a bandpass Purcell filter for qubit readout,'' 
\href{https://doi.org/10.1103/PhysRevA.92.012325}{\emph{Phys. Rev. A} \textbf{92}, 012325 (2015)}.

























\end{thebibliography}

\end{document}